\documentclass[a4paper,11pt]{article}
\pdfoutput=1 

\def\msusy{\ensuremath{M_\text{SUSY}}}
\def\mgut{\ensuremath{M_\text{GUT}}}
\def\mzpred{\ensuremath{M_Z(\text{pred})}}
\def\mzexp{\ensuremath{M_Z(\text{exp})}}

\usepackage{cancel}
\usepackage{jheppub} 
\usepackage[T1]{fontenc} 
\usepackage{overpic}

\usepackage{color}
\usepackage{amsthm}
\usepackage{graphicx}
\usepackage{slashed}
\usepackage{amssymb}

\def\slashchar#1{\setbox0=\hbox{$#1$}           
   \dimen0=\wd0                                 
   \setbox1=\hbox{/} \dimen1=\wd1               
   \ifdim\dimen0>\dimen1                        
      \rlap{\hbox to \dimen0{\hfil/\hfil}}#1 
   \else                                        
      \rlap{\hbox to \dimen1{\hfil$#1$\hfil}}/                                    \fi}

\DeclareMathOperator{\sgn}{sgn}

\title{\boldmath Investigating Multiple Solutions in the Constrained
  Minimal Supersymmetric Standard Model}

\author[a]{B C Allanach,}
\author[a,b]{Damien P. George} 
\author[c]{and Benjamin Nachman}

\affiliation[a]{DAMTP, CMS, University of Cambridge, Wilberforce Road,
  Cambridge, CB3 0HA, U.K.} 

\affiliation[b]{Cavendish Laboratory, University of Cambridge,
  JJ Thomson Avenue, Cambridge, CB3 0HE, U.K.}

\affiliation[c]{SLAC, Stanford University, 2575 Sand Hill Rd, Menlo Park,
  CA 94025, U.S.A.}

\emailAdd{B.C.Allanach@damtp.cam.ac.uk}
\emailAdd{dpg39@cam.ac.uk}
\emailAdd{bnachman@cern.ch}

\abstract{Recent work has shown that the Constrained Minimal
  Supersymmetric Standard Model (CMSSM) can possess several distinct solutions
  for certain values of its parameters. The extra 
  solutions were not previously found by public supersymmetric spectrum
  generators because fixed point iteration (the algorithm used by the
  generators) 
  is unstable in the neighbourhood of these solutions.
  The existence of the additional solutions calls into question 
  the robustness of exclusion limits derived from collider experiments and 
  cosmological observations
  upon the CMSSM, because limits were only placed on one of the
  solutions. Here, we map the CMSSM by exploring its multi-dimensional
  parameter space using the {\em shooting method}, which is not subject to 
  the stability issues which can plague fixed point iteration. We are able to
  find multiple solutions where in all 
  previous literature only one was found.
  The multiple solutions are of two distinct classes. One class, close
  to the border of bad electroweak symmetry breaking, is disfavoured by LEP2
  searches for neutralinos and charginos.
  The other class has sparticles that are heavy enough to evade the LEP2
  bounds.  
  Chargino masses may differ by up to around 10$\%$ between the different
  solutions, whereas other sparticle masses differ at the sub-percent level. 
  The prediction for the dark matter relic density can
  vary by a hundred percent or more between the different solutions, so
  analyses employing the dark matter constraint are incomplete without their
  inclusion.
}

\begin{document} 
\maketitle
\flushbottom

\section{Introduction}

It has been recently discovered that for a given set of input parameters for
the Constrained Minimal Supersymmetric Standard Model
(CMSSM)~\cite{Fayet:1976et,Fayet:1977yc,Farrar:1978xj,Fayet:1979sa,Dimopoulos:1981zb},
there can exist several distinct solutions for the particle
spectrum~\cite{Allanach:2013cda}.  This is because the renormalisation group
equations (RGEs) connecting the high- and low-scale observables form a boundary
value problem, which may admit multiple solutions.
A priori, this fact has important consequences for experimental collider
searches for 
new physics.  Previous analyses that placed bounds on specific regions of the
CMSSM parameter space used spectrum generators which missed the additional
solutions.  The published limits are therefore optimistic, as they ignore these
additional, potentially viable solutions.  We present here a study which shows
that this 
initial cause of concern is not necessary for many sparticle searches: the
phenomenology of the additional 
solutions are constrained by current and future experiments. However, there
are particular cases (notably, those involving charginos), where care is
needed and the additional solutions could provide a loop-hole to exclusion.

It was realised some years ago that the minimal supersymmetric standard model
has an LHC inverse problem~\cite{ArkaniHamed:2005px}. The problem is
essentially that LHC observables possess degeneracies such that the Lagrangian
of the MSSM cannot be uniquely determined from them. It was thought though,
that in strongly constrained set-ups like the CMSSM with only a few extra
parameters and many potential LHC observables, that there would be no inverse
problem (although even with fairly precise edge measurements, one could not
always discriminate between minimal gauge mediated supersymmetry breaking
patterns and the CMSSM~\cite{Allanach:2011ya}).  
Multiple solutions though, demonstrate that
even the CMSSM has its ambiguities and suffers from the {\em opposite}\/ of the
LHC inverse problem. While the LHC inverse problem has several different
parameter points leading to the same set of LHC observables, multiple
solutions have the {\em same}\/ CMSSM parameters leading to potentially {\em
  different}\/ LHC observables. 
One must be careful to be aware of the multiple solutions so as to not rule
out the 
model just because one is ignorant of an additional solution. The same caveat
potentially applies to other models of SUSY breaking. This does not
reintroduce the inverse problem, as sufficiently accurate observables will
discriminate between the different solutions, since they are physically distinct.

The CMSSM is a model of softly broken supersymmetry (SUSY) that imposes many
relations amongst the possible terms in the broken MSSM\@.  In particular, at
the Grand Unified Theory (GUT) scale, the scalar supersymmetric particles have
the same mass $m_0$, the gauge fermion supersymmetric particles have the mass
$M_{1/2}$ and the trilinear scalar couplings (Higgs-sfermion-sfermion) are
given by a new parameter $A_0$ multiplied by the corresponding Standard Model
Yukawa matrices.  Additionally, the parameter $\tan\beta$ specifies the
ratio of the two Higgs' vacuum expectation values at the SUSY breaking scale.
The final input to the CMSSM is the sign of the Higgsino mass term,
$\mathrm{sign}(\mu)$.  The value of $\mu$ is set by requiring the calculated
$Z^0$ mass is equal to the measured value (c.f.\ Eq.~\eqref{mzpred}).
The SUSY 
particle spectrum at any given scale is then determined by solving the
RGEs with boundary conditions at the three scales: GUT, SUSY breaking, and
electroweak.

Specifying constraints at more than one scale creates a boundary value problem
which, in general and in practice~\cite{Allanach:2013cda}, can admit multiple
solutions. This is different to the case where a single boundary condition
has multiple solutions itself (see, for example the case of mSUGRA in
Ref.~\cite{Drees:1991ab}, which gave up to three solutions for $\tan \beta$). 
The existence of multiple solutions to the CMSSM RGEs for a given sign of $\mu$
have evaded the standard SUSY spectrum generators such as {\tt
  SOFTSUSY}~\cite{Allanach:2001kg}, {\tt ISASUSY}~\cite{Baer:1993ae}, {\tt
  SPheno}~\cite{Porod:2003um}, and {\tt SUSPECT}~\cite{Djouadi:2002ze}.  The
standard procedure for calculating supersymmetric spectra by all the current
generators is fixed point iteration.  This algorithm is blind to the extra
solutions because they correspond to unstable fixed points~--- the standard
methods move away from them by necessity.  This fixed point iteration method
can be modified, and partially improved, by choosing to {\it scan}\/ over one of
the dimensions of the RGE and also relax one constraint.  For instance, one can
scan the value of $\mu$, relax the condition on $M_Z$, and then investigate the
locations in the one dimensional scan that give the correct prediction for
$\mzexp$. 
  
The more general method for finding all possible multiple solutions (at
least within the available numerical precision and computing time) is to
express the RGEs and the input data as an initial value problem. 
The Cauchy-Lipschitz theorem~\cite{CL} states that this initial value problem
has a unique solution provided the RGE trajectories are Lipschitz continuous
(they are unless poles are encountered). 
For the
CMSSM in the dominant third family approximation\footnote{We shall work in the
dominant third family approximation here, where the only Yukawa coupling
entries that are non-zero are those in the $(3,\ 3)$ position.}, one has 11
high-scale 
parameters to fix, and then runs the RGEs down 
to the lower scales to predict 11 quantities, most of which are Standard Model
observables.  The errors in the low-scale predictions are then used to
formulate a root-finding problem: find the high-scale parameters that give
negligible error for all predictions.  We shall show in this paper how to
implement this method using both the multidimensional Newton-Raphson and
Broyden root finders.
We then apply this general solver to well-studied slices of the CMSSM
parameter 
space, to identify regions with multiple solutions and systematically study
their 
phenomenology.

The paper is organised as follows.
In Sec.~\ref{sec:cmssm} we review the calculation of the spectrum of
the CMSSM, and how it is implemented in publicly available computer
programs.  Sec.~\ref{sec:rgemethods} discusses the instability of fixed
point iteration, and describes an alternative, the shooting method, which
can be used to explore the full solution space of a model defined at
multiple scales.  The details of our numerics are given in
Sec.~\ref{sec:numerics}.  In Sec.~\ref{sec:pheno} we present a map
of multiple solutions in the CMSSM (for a few representative points in
parameter space) and discuss the phenomenology of these solutions, giving
limits based on collider and astrophysical data.
Finally, Sec~\ref{sec:summary} summarises the work with a discussion on
the impact of the multiple solutions on experimental exclusions. 

\section{Spectrum calculation in the CMSSM}
\label{sec:cmssm}

To compute the observable spectrum of SUSY particles in the CMSSM one
must determine the MSSM parameters in the model by solving their RGEs with
certain boundary conditions.  The high-scale boundary conditions of the
CMSSM specify a theoretically motivated input on the soft supersymmetry-%
breaking mass terms in the Lagrangian, and the low-scale boundary conditions
match parameters to known data on Standard Model fermion masses and mixings,
gauge boson masses and gauge couplings. Over one hundred coupled, homogeneous,
non-linear ordinary differential equations evolve the couplings between the
different scales.  At the high scale there are a certain number of fixed
parameters (due to the GUT unification conditions) and the rest are initially
free but must be solved for, such that all low scale constraints are satisfied.
In this section we detail these boundary conditions and describe the existing
method used to numerically compute solutions.

Boundary conditions are specified at three scales: the low scales $M_Z$
and $\msusy$, and the high scale $\mgut$.  The latter two scales are
part of the set of the free parameters and are themselves determined
by the boundary conditions.  The conditions at these scales are
specified below, in terms of: the Yukawa couplings of the top, bottom,
and tau, {\em viz.} $h_t$, $h_b$, and $h_\tau$, respectively;
the 3 MSSM gauge couplings, $g_i$, $i \in \{1,2,3\}$; the various SUSY
breaking scalar masses, $m_{\varphi}$; the gaugino masses, $M_i$,
$i \in \{1,2,3\}$; and the SUSY breaking trilinear scalar couplings for
top, bottom, and tau, $A_t, A_b$ and $A_\tau$.  Also included are
constraints on the parameter $\mu$ appearing in the
superpotential\footnote{The circumflex indicates a superfield.}
$W \supset \mu \hat H_1 \hat H_2$, and constraints on $m_3^2$, the
parameter that mixes the two Higgs doublets in the potential $V \supset m_3^2
H_2 H_1$.  These latter two conditions (Eqs.~\eqref{mucond},\eqref{Bcond}
below) come from the 
minimisation 
of the Higgs potential with respect to the neutral components of  
$H_1$ and $H_2$.  In terms of these variables the boundary conditions
are
\begin{eqnarray}
\tan \beta(M_Z) &=& \tan \beta (\mbox{input}), \label{tanb} \\
h_t({M_Z}) &=& \frac{m_t(M_Z) \sqrt{2}}{v(M_Z) \sin \beta}, \qquad
h_{b,\tau}(M_Z) = \frac{m_{b,\tau}(M_Z) \sqrt{2}}{v(M_Z) \cos
  \beta}, \label{Yukawas}\\
v(M_Z) &=& 2 \sqrt{
  \frac{\mzexp^2 + \Pi_{ZZ}^T(M_Z)}
  {\frac{3}{5} g_1^2(M_Z) + g_2^2(M_Z)} },\label{vev}\\ 
g_1(M_Z)&=&g_1(\mbox{exp}), \qquad g_2(M_Z) = g_2(\mbox{exp}), \qquad g_3(M_Z) =
g_3(\mbox{exp}), \label{gaugeCouplings} \\
\msusy &=& \sqrt{m_{{\tilde t}_1}(\msusy) m_{{\tilde
      t}_2}(\msusy)}, \label{msusy} \\
\mu^2(\msusy) &=&  
\frac{m_{\bar{H}_1}^2(\msusy) -  m_{\bar{H}_2}^2(\msusy) \tan^2
  \beta(\msusy)}{\tan^2 \beta(\msusy) - 1} 
- \frac{1}{2} M_Z^2(\msusy),
\label{mucond} \\
  \tan \beta(\msusy) &=& \tan \frac{1}{2} \left[ \sin^{-1} 
    \left( \frac{2 m_3^2(\msusy)}{m_{\bar H_1}^2(\msusy) + m_{\bar
          H_2}^2(\msusy) + 
      2 \mu^2(\msusy)}\right)
\right],
 \label{Bcond} \\
g_1(\mgut) &=& g_2(\mgut), \label{unifc} \\
M_1(\mgut) &=& M_2(\mgut) = M_3(\mgut) = M_{1/2}, \label{mhalf} \\
m_{\tilde u}^2(\mgut) &=& m_{\tilde d}^2(\mgut) = m_{\tilde e}^2(\mgut) = m_{\tilde L}^2(\mgut) =
m_{\tilde Q}^2(\mgut) = m_0^2 I_3, \label{sparticle}
\\ m_{H_1}^2(\mgut) &=&
m_{H_2}^2(\mgut) =m_0^2, \label{mzero} \\ 
A_{\tilde u}(\mgut) &=& A_0 I_3, \qquad
A_{\tilde d}(\mgut) = A_0 I_3, \qquad
A_{\tilde e}(\mgut) = A_0 I_3. \label{trilinears}
\end{eqnarray}
The running parameters in Eqs.~\eqref{tanb}-\eqref{trilinears} are in
the modified dimensional reduction (DRED) scheme~\cite{Capper:1979ns}.
The `exp' denotes that the value derives from experimental data. 
We have labelled the input parameter $\tan \beta$ as
$\tan \beta(\mbox{input})$ to discriminate it from the predicted value run down
from the GUT scale.  The parameters
$m_{b,t,\tau}(M_Z)$ and $g_{1,2,3}(M_Z)$ are obtained from experimental
data, subtracting loops due 
to sparticles and Standard Model particles. The Standard Model electroweak
gauge couplings  
$g_1(\text{exp})$ and $g_2(\text{exp})$ are fixed by the Fermi constant
$G_F$ and the fine structure constant $\alpha$.
The values of $g_i(\text{exp})$ are corrected by 
one-loop corrections involving sparticles.
The parameter $v(M_Z)\approx 246$ GeV denotes
$\sqrt{v_1^2(M_Z)+v_2^2(M_Z)}$, where $v_1$ and $v_2$ are the vacuum
expectation values of the neutral components of the Higgs doublets $H_1$ and
$H_2$, respectively. 
The modified DRED $Z^0$ boson mass squared is fixed by
$M_Z^2(\msusy)=v^2(\msusy) \left(\frac{3}{5} g_1^2(\msusy) +
  g_2^2(\msusy)\right)/4$.  Furthermore,
$m_{\bar{H}_i}^2=m_{H_i}^2-t_i/v_i$  are fixed by the soft SUSY
breaking mass parameters for the Higgs fields $m_{H_i}^2$, 
$i \in \{1,2\}$, as well as by the tadpole
contributions $t_i$ coming from loops. 
These tadpole contributions have terms linear in $\mu(\msusy)$ as well as
terms that are logarithmic in it, and so Eq.~\eqref{mucond} is not a
simple quadratic equation for $\mu(\msusy)$.
The symbol $\Pi_{ZZ}^T(M_Z)$ denotes the MSSM self-energy correction to the $Z^0$ boson mass
which can be found in Ref.~\cite{Pierce:1996zz} and
$I_3$ is a 3$\times$3
matrix in family space. 
For further details, see the {\tt SOFTSUSY}~manual~\cite{Allanach:2001kg}. 

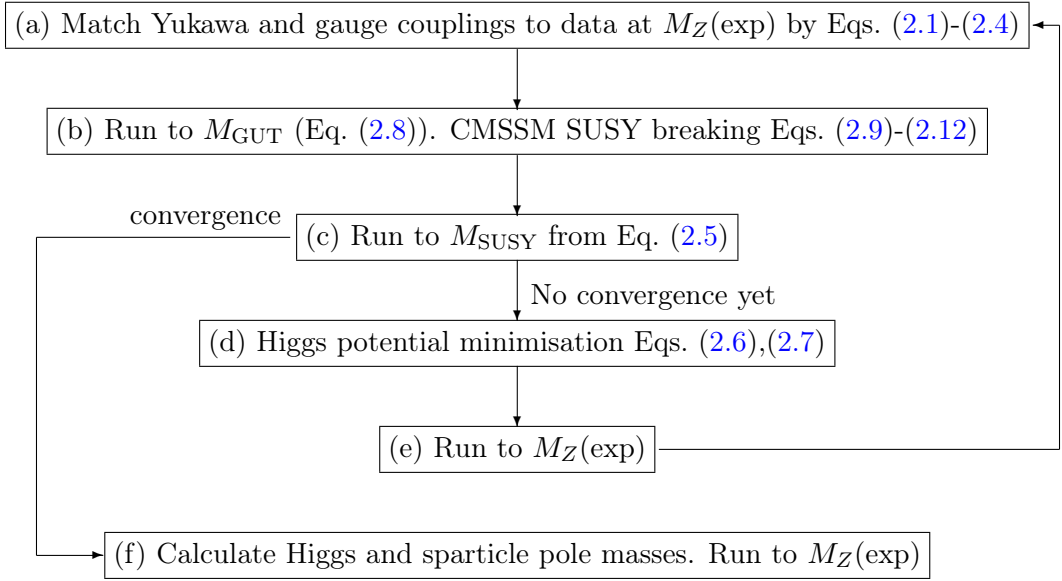
\begin{figure}\begin{center}
\begin{picture}(323,210)
\put(10,0){\makebox(280,10)[c]{\fbox{(f) Calculate Higgs and
      sparticle pole masses. Run to $\mzexp$}}}
\put(10,40){\makebox(280,10)[c]{\fbox{(e) Run to $\mzexp$}}}
\put(150,76.5){\vector(0,-1){23}}
\put(10,160){\makebox(280,10)[c]{\fbox{(b) Run to $\mgut$
      (Eq.~\eqref{unifc}). CMSSM SUSY breaking
      Eqs.~\eqref{mhalf}-\eqref{trilinears}}}} 
\put(150,116.5){\vector(0,-1){23}}
\put(10,80){\makebox(280,10)[c]{\fbox{(d) Higgs potential minimisation Eqs.~\eqref{mucond},\eqref{Bcond}}}}
\put(150,156){\vector(0,-1){23}}
\put(10,120){\makebox(280,10)[c]{\fbox{(c) Run to $\msusy$ from
      Eq.~\eqref{msusy}}}} 
\put(65,125){\line(-1,0){95}}
\put(-30,125){\line(0,-1){120}}
\put(-30,5){\vector(1,0){25}}
\put(5,130){convergence}
\put(155,100){No convergence yet}
\put(150,196){\vector(0,-1){23}}
\put(10,200){\makebox(280,10)[c]{\fbox{(a)
      Match Yukawa and gauge couplings to data at $\mzexp$ by Eqs.~\eqref{tanb}-\eqref{gaugeCouplings}}}}  
\put(203,45){\line(1,0){150}}
\put(353,45){\line(0,1){160}}
\put(353,205){\vector(-1,0){10}}
\end{picture}
\caption{\label{fig:algorithm}
  Fixed point iteration algorithm used by current publicly available
  SUSY spectrum calculators to calculate the SUSY
  spectrum. The initial
step is the uppermost one.}\end{center}\end{figure}

Spectrum calculators for the MSSM that
are currently in the public domain, namely {\tt
  ISASUSY}~\cite{Baer:1993ae}, {\tt 
  SOFTSUSY}~\cite{Allanach:2001kg}, {\tt   SPheno}~\cite{Porod:2003um}, {\tt
  SUSEFLAV}~\cite{Chowdhury:2011zr} and {\tt SUSPECT}~\cite{Djouadi:2002ze},
use fixed point iteration to find a self-consistent solution to
Eqs.~\eqref{tanb}-\eqref{trilinears} (or equations very similar to them).
The particular algorithm used by {\tt SOFTSUSY}, for example, is shown in
Fig.~\ref{fig:algorithm}. The algorithm begins at $M_Z$ by guessing values of
all MSSM parameters at that scale. Threshold corrections to gauge and Yukawa
couplings are calculated, and they are then matched to data including
sparticle loop corrections. The MSSM parameters are then run to $\mgut$, where 
the SUSY breaking equations are imposed. Running down to $\msusy$, parameters
are fixed such that the Higgs potential gives the desired electroweak
minimum. The MSSM parameters are then run back down to $M_Z$, where we return
to point (a). This loop is continued until the MSSM parameters converge to within
some fixed tolerance: i.e.\ running around the loop no longer changes them. 
When the algorithm finishes (point (c) in the figure), pole masses of Higgs'
and sparticles are calculated. 

This iterative algorithm is an example of fixed point iteration,
where one attempts to find a set of values that remain invariant
under application of a certain function.  In the case at hand the
values are the parameters of the MSSM and the function (actually a set
of functions, one for each parameter) is one iteration of the loop in
Fig.~\ref{fig:algorithm}.  If a solution to the system of MSSM RGEs is
found then it is considered the one and only solution.   Such an assumption
is na\"ive, and it was shown in Ref.~\cite{Allanach:2013cda} that, due to
the boundary value nature of the system, multiple solutions can, and do
in fact, exist, depending upon CMSSM parameters.  
Amongst a set of multiple solutions, many of the $M_{GUT}$ scale parameters
(the ones that are set by the lower scale boundary conditions) 
are different, and so are many low-scale observables.  Some are more
different than others, however. The Higgs potential parameters
$\mu$ and $m_3^2$ can be quite different between the different solutions. 
$\mu$ changes the stop, sbottom and stau mixing. This then changes the top,
bottom and tau Yukawa couplings at {\em leading log} order (i.e.\ first order
in $1/(16 \pi^2)\log (\mgut/M_Z)$), because the
measurement of the top, bottom or tau mass would include stop, sbottom or stau
radiative corrections, respectively\footnote{Neutralino and chargino
masses and mixings are also affected, changing various radiative corrections
(for example to the extracted electroweak gauge couplings). Thus, the
electroweak gauge couplings also change to leading log order.}. The fact
that these couplings change then changes the RGE 
trajectories of most of the soft masses. Strictly speaking, these are only thus
changed beyond leading log order (since it's a loop effect induced by 
an effect that is already leading log), but the 
large separation of scales between the weak
and supersymmetry breaking scale can enhance the higher order logs. 

Whilst fixed point iteration
can in some cases use different starting conditions to find multiple solutions,
we shall demonstrate in the next section that there are certain solutions which
the algorithm can never converge on.  We shall also show that such solutions
are sometimes 
important in the context of CMSSM phenomenology.

\section{RGE Solution Methods and Stability}
\label{sec:rgemethods}

In this section we first describe the potential non-robustness of fixed
point iteration due to instabilities, and then detail a method of solving
the multi-boundary problem that does not suffer from stability issues. 

\subsection{Fixed Point Iteration}
\label{sec:fpi}
As mentioned in Sec.~\ref{sec:cmssm}, current publicly available SUSY spectrum
generators use {\em fixed point iteration}\/ (FPI) to solve the multi-boundary
problem relevant to the CMSSM\@.  FPI operates by making a guess for the unknown
high-scale parameters and repeatedly runs the RGEs down and up, applying
the constraints until the solution no longer changes above some prescribed
level.  It has the distinct advantage of being very quick, as often only a few
iterations are required in order to achieve convergence to some desired numerical
accuracy.  However, FPI generically suffers from stability issues: for a given
point in parameter space, certain solutions may be unstable with respect to FPI,
and so the algorithm will never converge on them.
Here we review, following Ref.~\cite{fpiStab}, the stability of FPI, first in
the 
simple toy one-dimensional case, then generalising to the multi-dimensional case
which is the case relevant for the CMSSM. 

In one-dimension, FPI can be used to solve
\begin{equation}
x_*=f(x_*), \label{eq1}
\end{equation}
i.e.\ $x_*$ represents a fixed point of the function $f(x)$.
For FPI, we start with a guess $x_1$ of the parameter, and generate
(hopefully better) successive approximations by
\begin{equation}
x_{n+1}=f(x_n) \Leftrightarrow x_* - (x_* - x_{n+1}) = f(x_* - (x_* - x_n)).
\end{equation}
Taylor expanding the right hand side around $x_*$, we have, to first order
in $x_*-x_n$,
\begin{equation}
\cancel{x_*}-(x_*-x_{n+1})=\cancel{f(x_*)}-(x_*-x_n) \frac{df(x_*)}{dx}\Leftrightarrow 
(x_* - x_{n+1}) = (x_*-x_n) \frac{df(x_*)}{dx}.
\end{equation}
Stable convergence of FPI requires that, once $x_n$ is sufficiently close to
$x_*$, successive approximations are always closer: $| x_*- x_{n+1} | < |x_* - x_n|$.
Therefore,
\begin{equation}
\left| \frac{df(x_*)}{dx} \right| < 1.
\end{equation}
Conversely, if $\left| \frac{df(x_*)}{dx} \right| > 1$, then FPI will be
unstable around the solution.

In our multi-dimensional case,
${\bf x}$ represents MSSM couplings and masses, and ${\bf f}({\bf x})$
is the multi-dimensional function that represents what happens to ${\bf x}$
after running up to the high scale, applying boundary conditions,
running back down and 
applying the other boundary conditions, i.e.\ one iteration: the loop
shown in Fig.~\ref{fig:algorithm}.
In the following, bold capitals represent matrices, whereas bold font lower
case letters are vectors. We now have $N$ fixed point equations 
\begin{equation}
{\bf x_*}={\bf f}({\bf x_*}).
\end{equation}
Proceeding similarly as we did in the one-dimensional case, we solve this system
with successive approximations ${\bf x_n}$, where
\begin{equation}
{\bf x}_{n+1}={\bf f}({\bf x}_n) \Leftrightarrow {\bf x_*} - ({\bf x_*} - {\bf
  x}_{n+1}) = {\bf f}({\bf x_*} - ({\bf x_*} - {\bf x}_n)). 
\end{equation}
Taylor expanding the right hand side around ${\bf x_*}$, we
have, to first order in $({\bf x}-{\bf x}_n)$, 
\begin{equation}
\cancel{{\bf x_*}}-({\bf x_*}-{\bf x}_{n+1})=\cancel{{\bf f}({\bf
    x_*})}-\left[({\bf 
  x_*}-{\bf x}_n)^T\cdot {\bf \nabla}\right] {\bf f}|_{\bf x^*}\Leftrightarrow 
({\bf x_*} - {\bf x}_{n+1}) = {\bf D} ({\bf x_*}-{\bf x}_n), \label{fo}
\end{equation}
where ${\bf D}$ is the {\em Jacobian matrix}\/ of $\bf f$, evaluated at $\bf
x_*$. 
Defining the {\em matrix norm} $|| {\bf D} ||$ of ${\bf D}$,
\begin{eqnarray}
||{\bf D}|| &\equiv&
\mbox{max}_{\bf x :\ ||{\bf x}||=1} \left\{\ || {\bf D} {\bf x} ||\  \right\}
= \mbox{max}_{\bf x :\ {\bf x} \neq 0} \left\{ \frac{|| {\bf D} {\bf x} ||}{||
  {\bf x} ||}  \right\}\nonumber \\
\Rightarrow|| {\bf D} ({\bf x}-{\bf x}_n)  || &\leq &  || {\bf D} || \
|| ({\bf x}-{\bf x}_n) ||.
\end{eqnarray}
Then Eq.~\eqref{fo} yields
\begin{equation}
|| {\bf D} || < 1 \Rightarrow || {\bf x_*}-{\bf x}_{n+1} || < || {\bf x_*}-{\bf
  x}_{n} || \label{mstab}
\end{equation}
as a stability condition. Conversely, from Eq.~\eqref{fo}, we have 
\begin{equation}
\mbox{min}_{\bf x: ||x||=1}\left\{ || {\bf D x} || \right\} > 1 \Rightarrow
|| {\bf x_*} - {\bf x_{n+1}} || > || {\bf x_*} - {\bf x_n}
||. \label{instability} 
\end{equation}

If Eq.~\eqref{instability} is satisfied, the solution is clearly unstable to
FPI\footnote{Note that there could also be solutions for which neither
  Eq.~\eqref{mstab} nor Eq.~\eqref{instability} are true. For such solutions,
  it is unclear 
  what the stability properties would be.}. 
Depending upon CMSSM parameters then, there may be solutions to the RGEs 
that have not been found by publicly available spectrum calculators,
all of which rely on FPI\@. This was demonstrated to be true in
Ref.~\cite{Allanach:2013cda}, 
where one of the boundary conditions (Eq.~\eqref{mucond}) was inverted:
$\mu(\msusy)$ was fixed, and the equation was used instead to predict $M_Z$. 
It was demonstrated that some points in CMSSM parameter space had {\em
  several different}\/ values of $\mu(\msusy)$ that fit the empirically measured
central value for $M_Z$ i.e.\ $\mzexp$. These constituted multiple
solutions of the RGEs for the same CMSSM point. Running FPI on some of the
additional solutions showed that they 
were unstable, with parameters getting successively further from the solution
in question.

\subsection{The Shooting Method}

The {\em shooting method}~\cite{numRec} provides an alternative method to FPI of
solving a multi-boundary problem. Instead of running the RGEs up and down, one
runs only in one direction, by first applying the boundary conditions relevant
for the starting scale and guessing any parameters that are not
fixed by these boundary conditions. It is convenient for us to start at
$\mgut$, applying the boundary conditions found in
Eqs.~\eqref{unifc}-\eqref{trilinears}. When one runs to the other scale (in our
case $\msusy$ then the weak scale), in general the other boundary conditions
are not 
exactly satisfied. One keeps on adjusting the values of the guessed GUT scale
parameters until, after running down, all of the $\msusy$ and weak-scale
boundary conditions are satisfied to some prescribed numerical accuracy.

We now face the fact that to truly have a chance of numerically finding all of
the solutions, we must hunt in eleven dimensions for each CMSSM point.  That is,
for each choice of $m_0$, $M_{1/2}$, $A_0$, $\tan \beta$ and $\sgn(\mu)$,
specifying 
a point in the CMSSM parameter space, there are 11 high-scale parameters (to be
detailed below) which must be determined such that the low-scale constraints are
satisfied.  The overall strategy is to start at the high scale $\mgut$,
input the 
MSSM parameters $y_i$, consistent with $\mgut$-scale CMSSM boundary
conditions, 
and evolve the RGEs down to $\msusy$ where the electroweak symmetry breaking
boundary conditions should be met, then to 
the weak scale, where various
weak-scale boundary conditions should be met. 

The MSSM renormalisation group equations may be phrased in general as
\begin{equation}
\frac{d y_i}{d t} = h_i(y_1,\ y_2,\ \ldots ,\ y_N), \label{RGEs}
\end{equation}
where, $i \in \{1,2, \ldots, 109\}$ and $t=\ln (Q/\textrm{GeV})$, $Q$ being the 
modified dimensional reduction scheme~\cite{Capper:1979ns} renormalisation
scale, in units of GeV.
The $h_i$ are in general non-linear functions of the
relevant MSSM parameters $y_i$. Here, for a given 
CMSSM point, we have the dimensionless $\sim {\mathcal O}(1)$ input parameters
displayed in Table~\ref{tab:inputs}.
\begin{table}
\begin{center}
\begin{tabular}{|ll|} \hline 
$V_1=\tan \beta(\mgut)/10$   &  $V_7=Y_t(\mgut)$\\ 
$V_2=\ln (\mgut/\mbox{1 GeV})$ & $V_8=Y_b(\mgut)$  \\
$V_3=g_1(\mgut)$ & $V_9=Y_\tau(\mgut)$\\
$V_4=g_3(\mgut)$ & $V_{10}=v(\mgut)/1000\mbox{~GeV}$\\
$V_5=[\mu(\mgut)/\mbox{1000 GeV}]^2$ & $V_{11}=\msusy(\mgut)/1000\mbox{~GeV}$\\
$V_6=m_3^2(\mgut)/10^6\mbox{~GeV}^2$ & \\
\hline\end{tabular}
\end{center}
\caption{\label{tab:inputs} Dimensionless order one input variables for each
  CMSSM point.} 
\end{table}
Eqs.~\eqref{unifc}-\eqref{trilinears} are applied upon all of the soft SUSY
breaking terms, and the MSSM couplings and parameters are evolved from
$Q/\textrm{GeV}=\exp(V_2)$ to $Q/\textrm{GeV}=1000 V_{11}$.

We wish to find values of the $V_i$ such that the boundary conditions at
$\msusy$ and $M_Z$ are satisfied: 
\begin{equation}
f_i(V_1,\ V_2,\ \ldots,\ V_{11})=0, \label{cond}
\end{equation}
where the definitions of the $f_i$ are shown in Table~\ref{tab:bcs}.
\begin{table}
\begin{center}
\begin{tabular}{|ll|ll|} \hline 
$f_1=\mzpred^2/\mzexp^2-1$ & $f_2=\tan
\beta(\msusy)/\tan \beta^{\textrm{pred}}(\msusy)-1$ \\
$f_3=\msusy/\msusy^{\text{pred}}-1$ & $f_4=Y_t^{\textrm{pred}}(\mzexp))
/ Y_t(\mzexp)) -1$ \\
$f_5=Y_b^{\text{pred}}(\mzexp) / Y_b(\mzexp) -1$ &
$f_6=Y_\tau^{\text{pred}}(\mzexp) / Y_\tau(\mzexp) -1$ \\
$f_7=g_1^\text{pred}(\mzexp) / g_1(\mzexp)-1$ & 
$f_8=g_2^\text{pred}(\mzexp) / g_2(\mzexp)-1$ \\
$f_9=g_3^\text{pred}(\mzexp) / g_3(\mzexp)-1$ &
$f_{10}=v(\mzexp)/v^\text{pred}(\mzexp)-1$ \\
$f_{11}=\tan \beta(\mzexp) / \tan \beta(\textrm{input}) - 1$ & \\
\hline\end{tabular}
\end{center}
\caption{\label{tab:bcs} Output variables encoding the target boundary
  conditions. The various quantities are precisely defined in
  Appendix~\protect\ref{sec:defs}. Note that $f_{11}$ measures how far we are
  away from a desired input value of $\tan \beta$(input), whereas $f_2$
  measures how far we are away from one of the Higgs potential minimisation
  conditions, i.e.\ Eq.~\eqref{Bcond}.} 
\end{table}
Thus, we must adjust the 11 $V_i$ in order to satisfy Eq.~\eqref{cond},
providing a valid solution to the boundary conditions and RGEs (if such
a solution exists).  We can therefore view the system as 11 simultaneous
non-linear equations in the 11 $V_i$.  We
solve them by two different methods, depending upon the context: Broyden's
method~\cite{broyden} and 
a globally convergent Newton-Raphson method, which are both used to
find roots of a multi-dimensional function.
Here, we shall sketch the Newton-Raphson method that we employ; for more
detail on it, or for information about Broyden's method, see Ref.~\cite{numRec}.
We start with some guess for the initial parameters $V_i$, and calculate
the Jacobian matrix $J_{ij}=\partial f_i / \partial V_j$ approximately by
finite differences:\footnote{We give here the forward-difference approximation.
We have also tried the symmetric-difference approximation, but it runs at
almost half the speed due to the additional evaluations of $f_i$, and does
not give noticeably improved convergence or accuracy.}
\begin{equation}
J_{ij} \approx \frac{f_i(V_1,\ V_2,\ \ldots, V_j + \Delta V_j,\ V_{j+1},\ \ldots)
  - f_i(V_i)}{\Delta V_j}
\end{equation}
for small enough $\Delta V_j$. Since
\begin{equation}
f_i(V_j + \Delta V_j) = f_i(V_j)  + \sum_j J_{ij} \Delta V_j +
\mathcal{O}((\Delta V_j)^2), \label{fdv}
\end{equation}
we solve $\sum_j J_{ij} \Delta V_j=-f_i$ for $\Delta V_j$, 
which solves Eq.~\eqref{fdv} to leading order for $f_i(V_j+\Delta V_j)=0$.
Adding $\Delta V_j$ onto $V_j$
should therefore take us closer to a solution of Eq.~\eqref{cond}. In practice, 
we require the step to be a {\em descent direction}, i.e.\ to first order
\begin{equation}
 \frac12 \sum_j\frac{\partial (\sum_i f_i f_i)}{\partial V_j} \Delta V_j = 
\sum_{ijk} -f_i J_{ij} J_{jk}^{-1} f_k = -\sum_i f_i f_i < 0. \label{descent}
\end{equation}
If Eq.~\eqref{descent} is not satisfied in practice because we are too 
far from the minimum of $f_i f_i$ and the second order terms in
Eq.~\eqref{fdv} give a sizeable correction, the algorithm backtracks until we
find 
an acceptable step, as described in more detail in Ref.~\cite{numRec}.
These updates are
iterated until Eq.~\eqref{cond} is satisfied to a given numerical accuracy
(specified here to be $10^{-5})$.

Unlike fixed point iteration, the shooting method, implemented using
either Broyden or Newton-Raphson for the root finding, does not rely on any
restrictions 
on the size of any derivatives in the vicinity of a solution. It is therefore
able in principle to find a solution that is unstable to fixed point
iteration. For given initial values of the $V_i$, one solution (which has some
`catchment volume' in $\{ V_i \}$ where the Broyden or
Newton-Raphson method 
will converge to it, see Fig.~\ref{fig:catchment}) can be
found. The idea then is to run with several randomly chosen $V_i$ in order to
have the possibility of finding the usual fixed-point iteration solutions, the
solutions found in Ref.~\cite{Allanach:2013cda} by scanning $\mu(\msusy)$, and
any other solutions as well. We have
the problem that, for a given number of initial $V_i$, we can never guarantee
that there are no other solutions other than those we have found. This is a
general problem though in any numerical scan, and, failing some analytical
breakthrough,  there does not appear to
be much we can do about it. 

\section{Numerical Procedure}
\label{sec:numerics}

We use a modified version of {\tt SOFTSUSY3.3.6} that implements the shooting
method as described above. 
For a given CMSSM point, first we find a solution
using the usual FPI method\footnote{Sometimes such a solution is not
  physically viable, but we still use it as a starting point.}. 
We then randomly perturb around this point and run the Broyden or
Newton-Raphson root-finding algorithm in order to obtain a solution. 
The GUT scale parameters $V_i$ of this solution are checked
against those corresponding to any solutions previously found. If any of the
$V_i$ differs by at least 0.001, it is counted as an additional solution and
saved. If the $V_i$ are all within 0.001, we save the solution with the
smallest value of $\textrm{max}_i |f_i|$, since this represents the best
numerical approximation to solving the boundary conditions. 
We repeat 
this procedure 2000 times with different random starting GUT scale initial
parameters. It is hoped that 2000 times is sufficient to find all of the
solutions, but we can never be sure that this is the case. Our results are
usually phrased as scans
across the CMSSM parameters, so statistical noise will be seen in any
derived plots. 
Our random perturbations are based on draws from a Gaussian distribution
centred on zero of
width 1, which we denote $G_i(1)$ for each independent draw. Each of the $V_i$
are changed with a probability of 0.5.
If perturbed, the $V_i$  are multiplied in turn by $|1 + G_i(1) \times a_i|$
(where we must pick values for the real coefficients $a_i$), with the
exception of $V_6=[m_3^2(\mgut)/10^6 \textrm{~GeV}^2]$, which is 
instead multiplied by $G_i(1) \times a_6$. $V_6$ is singled out for different
treatment because we wish to consider either sign for its changed value.

The larger the $a_i$, the less likely
it is that {\tt SOFTSUSY} will be able to return a physical solution
(unphysical solutions could be due to the predicted
presence of, for example, tachyons).  On the other hand, larger
$a_i$ explores a larger volume of MSSM parameter space within which
extra solutions could be found.
Choosing all $a_i$ of order one gives an unacceptably small efficiency, so as
to make finding additional solutions extremely unlikely with 2000 `shots'. 
Indeed, trying to find any solution starting from a completely random starting
point proved to be impossible: the efficiencies were much too small to
find it in 2000 shots.

At tree-level, the $M_Z$ boundary conditions on the Yukawa and gauge couplings
are specified unambiguously by
Eqs.~\eqref{tanb}-\eqref{gaugeCouplings} and experimental data. The quantities
$\mu(\msusy)$ 
and $m_3^2(\msusy)$ as derived from electroweak symmetry breaking can differ
between multiple solutions, which in turn leads to electroweak one-loop
corrections to sparticle threshold corrections to the gauge and Yukawa
corrections. Some parameters $a_i$ correspond to 
boundary conditions that are fixed at tree-level, but may differ
only in electroweak loop corrections. 
These $a_i$ we set to 
have a small value, whereas those $a_i$ corresponding to boundary conditions
that may differ already at tree-level we set to have a larger value. 
By examining the efficiencies of finding solutions,
we find, by trial and error, that the values $a_i=\{0.03$, $0.1$, $0.03$,
$0.03$, $1.0$, $1.0$, $0.03$, $0.03$,
$0.03$, $0.03$, $0.1\}$ give
a reasonably high probability of finding the
multiple solutions that we already know about, while
retaining an efficiency (defined as the fraction of random
perturbations resulting in a physical solution) that is not too small. 

The {\tt SOFTSUSY} {\tt TOLERANCE} parameter fixes how accurately the RGEs are
solved.\  
max$_i|f_i|$ was also required to be less than {\tt
  TOLERANCE} within 40 iterations of the Newton-Raphson algorithm (or
Broyden's algorithm). We found that with higher values of {\tt TOLERANCE}, we
would obtain higher (and spurious) values of  the number of multiple solutions
at a CMSSM point: decreasing {\tt TOLERANCE} further meant that some of the
apparently different solutions were not good solutions after all, and some of
them were actually inaccurately evaluated versions of the same solution. 
{\tt TOLERANCE} was set to be 10$^{-5}$ for our results below. We have checked 
for many particular points with multiple solutions that reducing it further
did not eliminate any such solutions (reducing {\tt TOLERANCE} further
resulted in a higher CPU time cost). We found that the statistical efficiency
of our results did depend somewhat on whether we had implemented Broyden's
algorithm, or Newton-Raphson, and furthermore on the precise implementation. Where there
was a difference, we used the algorithm which found more multiple solutions, 
or which displayed less statistical noise in the results. 

\section{Regions of Multiple Solutions}
\label{sec:pheno}

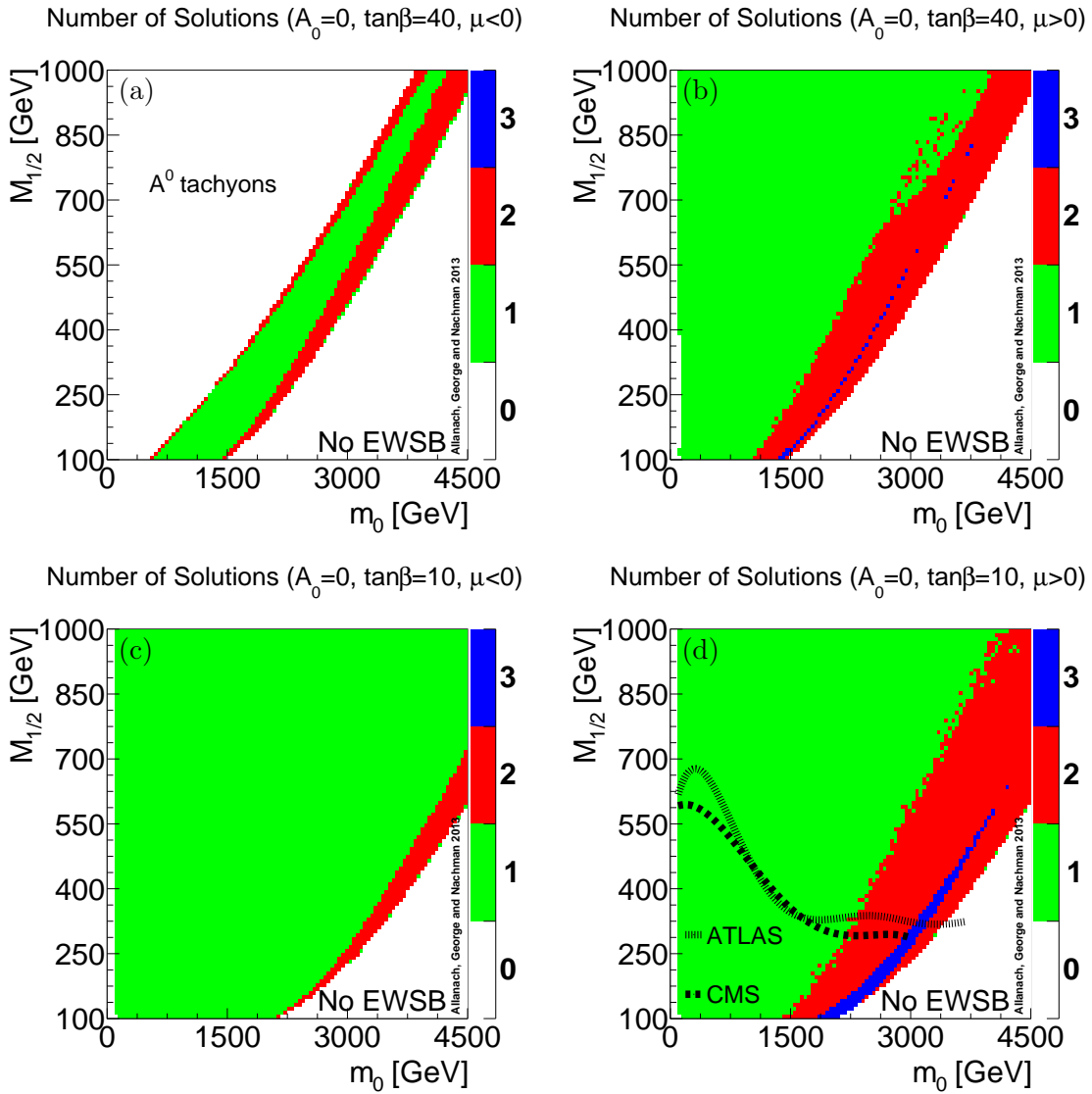
\begin{figure}
\begin{center}
\unitlength=\textwidth
\begin{picture}(1,0.8)
\put(0.0,0.5){\includegraphics[width=0.5 \textwidth]{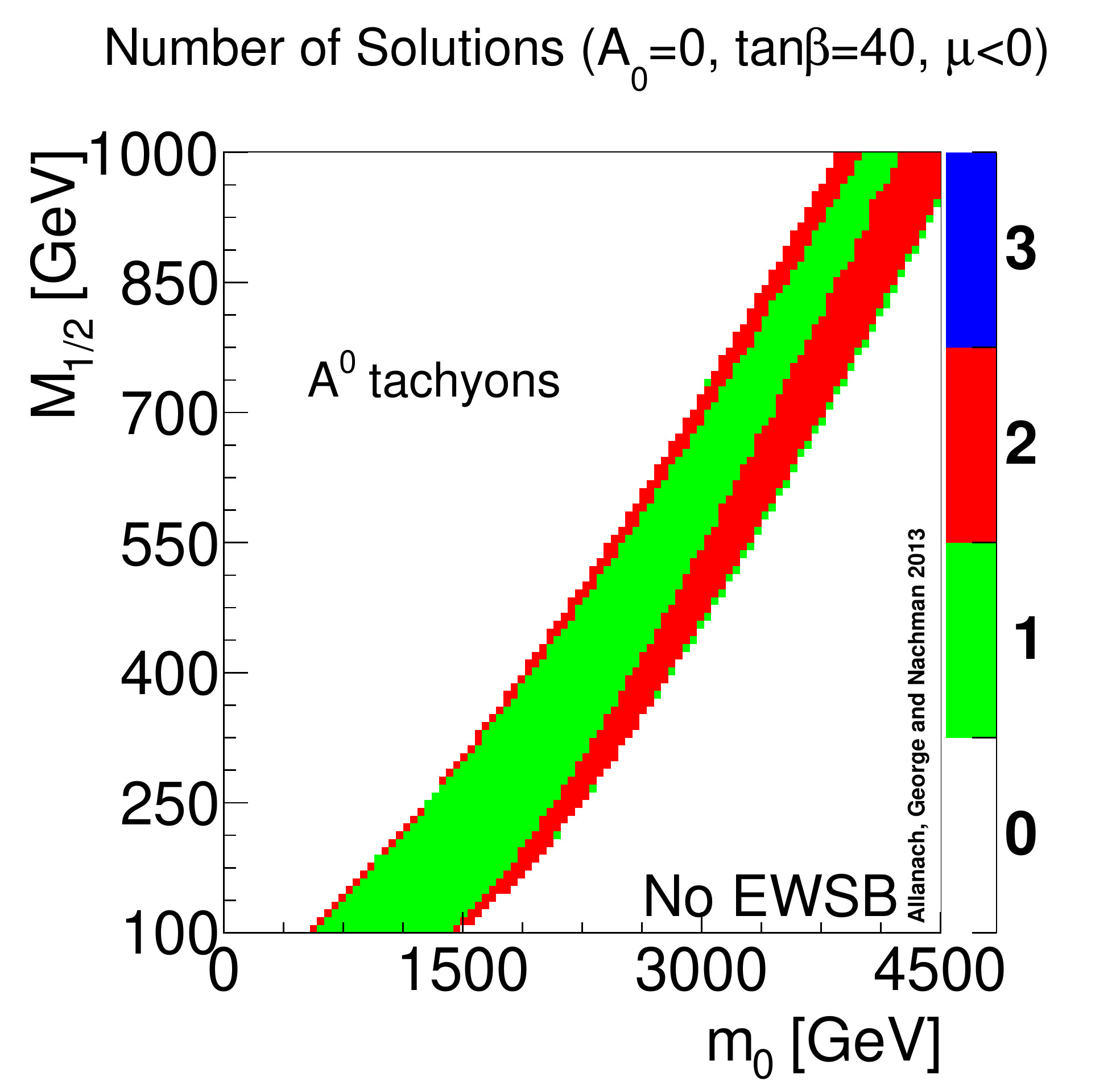}}
\put(0.5,0.5){\includegraphics[width=0.5\textwidth]{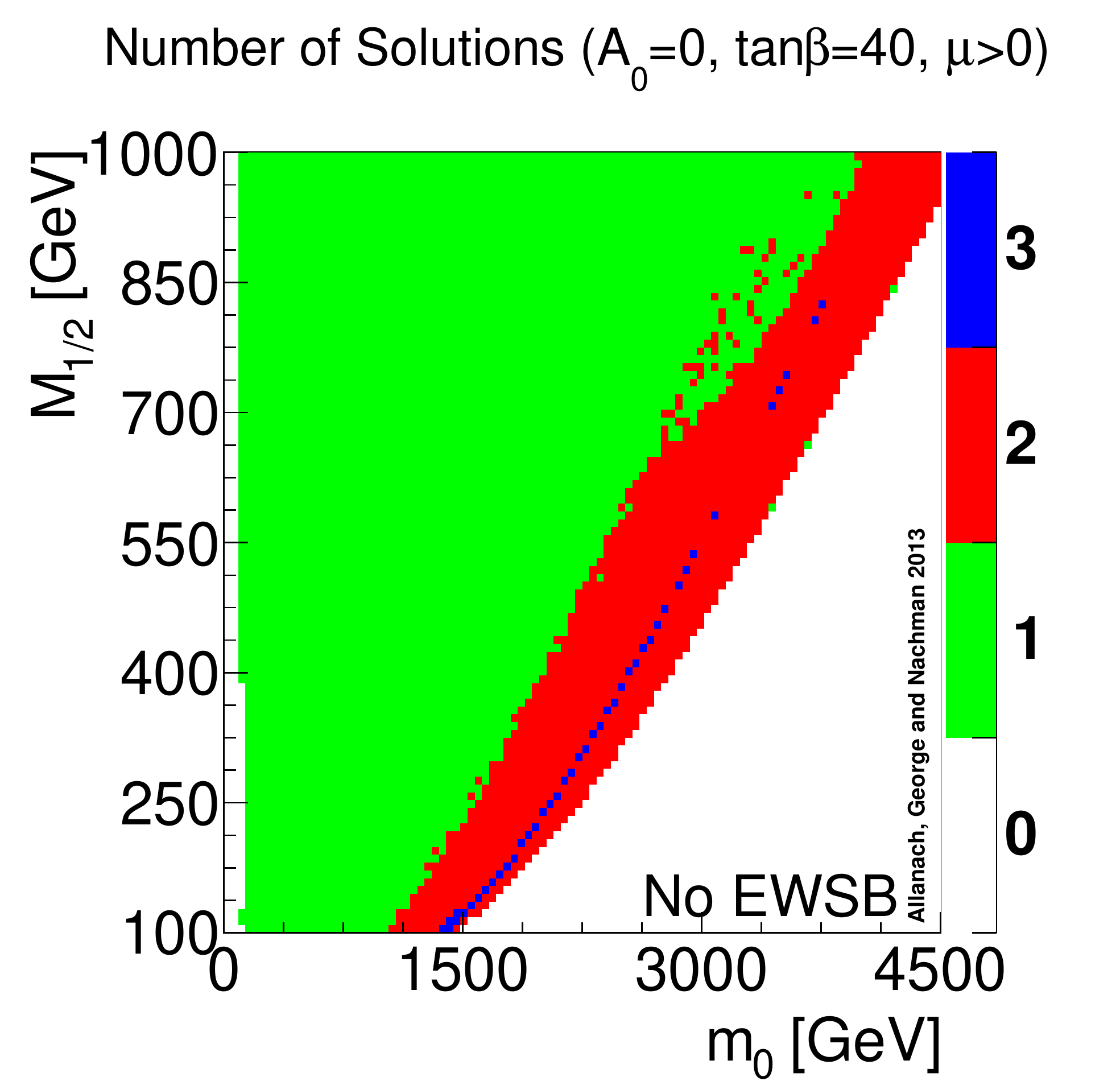}}
\put(0.0,0){\includegraphics[width=0.5\textwidth]{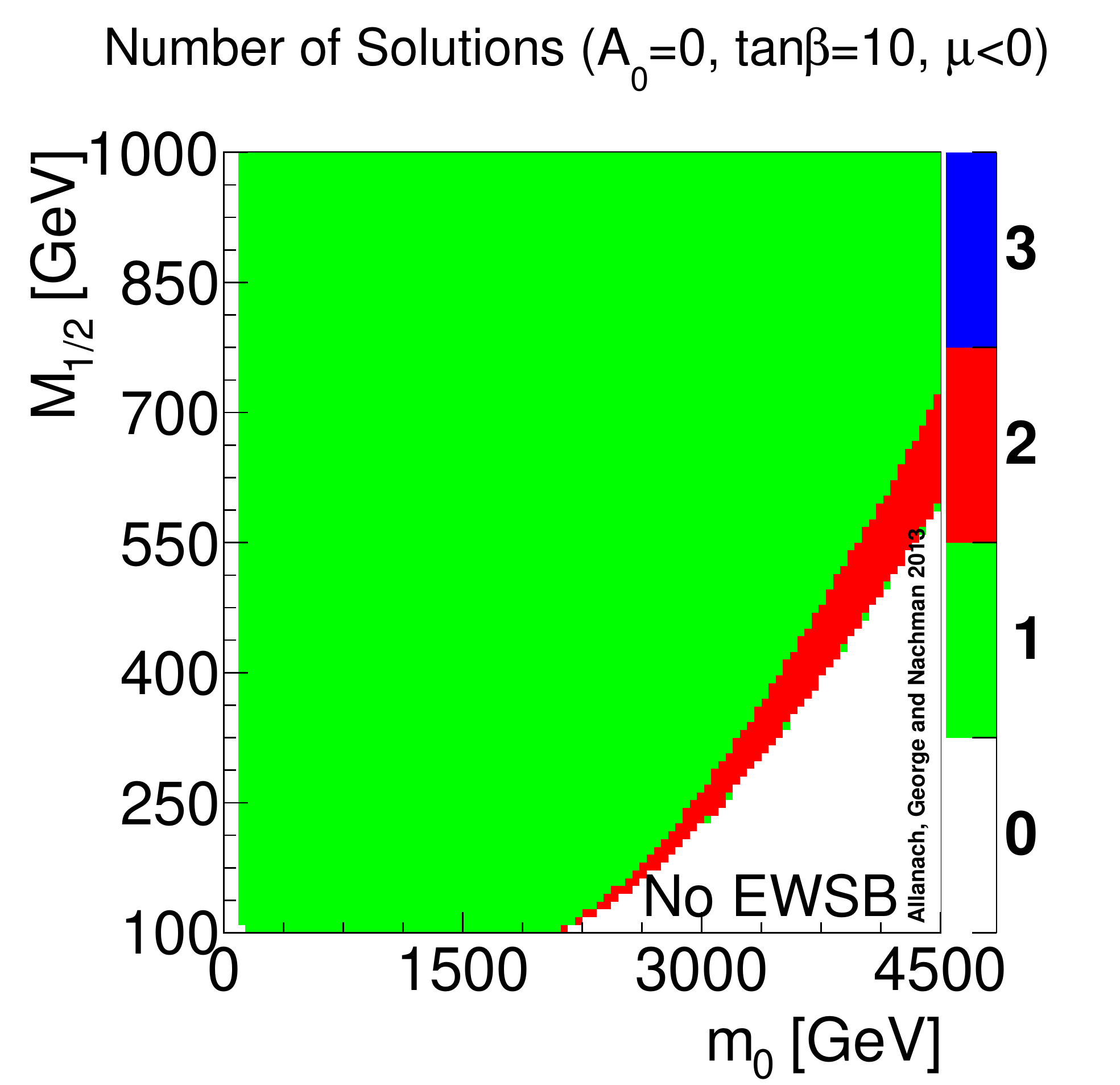}}
\put(0.5,0){\includegraphics[width=0.5\textwidth]{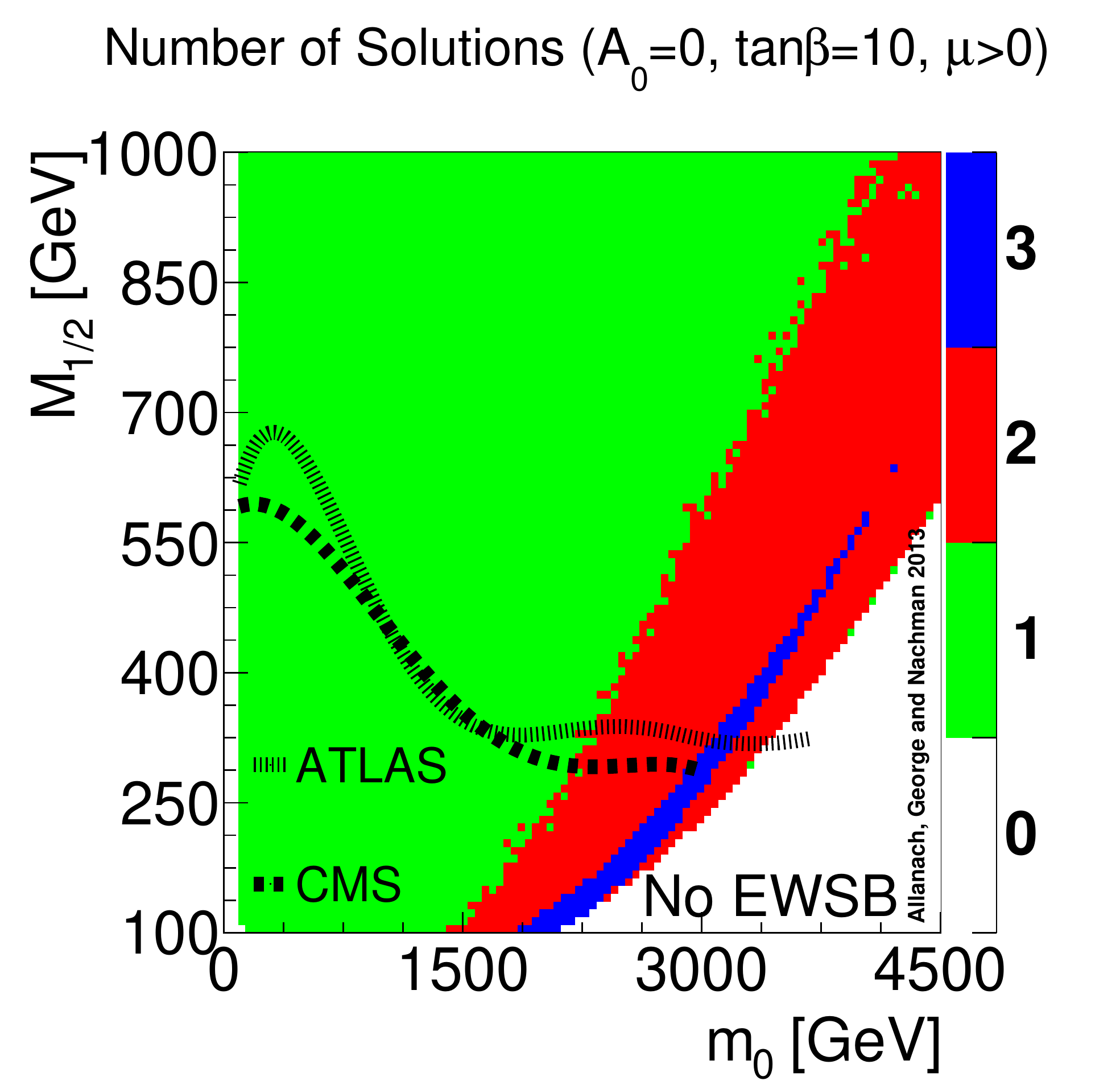}}
\put(0.11,0.89){(a)}
\put(0.61,0.89){(b)}
\put(0.11,0.39){(c)}
\put(0.61,0.39){(d)}
\end{picture}
\end{center}
\caption{\label{fig:numberofsolutions} Number of solutions in the CMSSM as
  shown as the background colour and labelled in the key on 
  the right-hand side of each plot for $A_0=0$ and different values of
  $\tan \beta$. White
  regions have no solutions for the reasons labelled: `No EWSB' denotes a
  region where there is no acceptable electroweak minimum of the Higgs
  potential. In (d), the lines display 95$\%$ exclusion contours
  from ATLAS \cite{Aad:2012fqa} and CMS~\cite{CMSsusy2012} jets plus
  missing transverse momentum searches. The region below
  each contour is excluded.}
\end{figure}
We first provide a map of CMSSM multiple solutions, as determined by the
multi-dimensional shooting method
outlined above using Broyden's method. First, we show the results of a scan of
$m_0$ and $M_{1/2}$ for $A_0=0$ and $\tan
\beta=40$ (the plane investigated most thoroughly in
Ref.~\cite{Allanach:2013cda}).
We see that for for $\mu<0$, there are two disjoint regions with two
solutions in Fig.~\ref{fig:numberofsolutions}a. In
Fig.~\ref{fig:numberofsolutions}b, we see that for $\tan \beta=40, A_0=0$ and
$\mu>0$ that
there is a region at high $m_0$ with multiple solutions near the boundary of
successful electroweak symmetry breaking. A line of three multiple solutions exists
within this band. Remembering that our method is stochastic in nature, we are
not surprised to see that this line is somewhat broken: we expect that, with
more statistics, the gaps would be filled in (expecting that the multiple
solutions are continuously connected in parameter space). Also, we see that to
the top 
left hand edge of the multiple solution region, there are scattered points. We
suspect too that here it is more difficult to find the multiple solutions, and
that were we to have more statistics (i.e.\ run for a higher number of
`shots'), we would fill in this area. We also see multiple solutions near the
electroweak symmetry breaking boundary at high $m_0$ for $\mu<0$ and $\tan
\beta=10$ in Fig.~\ref{fig:numberofsolutions}c and for $\mu>0$ and $\tan
\beta=10$ in Fig.~\ref{fig:numberofsolutions}d.
In Fig.~\ref{fig:numberofsolutions}d, we display the 95$\%$ confidence level
exclusion contour for ATLAS~\cite{Aad:2012fqa} and CMS~\cite{CMSsusy2012} searches for events with hard jets and large
missing transverse momentum in
5 fb$^{-1}$ of $\sqrt{s}=7$ TeV LHC data. We see that
this exclusion region covers much of the multiple solutions region. However,
the exclusion region was only calculated using the standard CMSSM solution, so 
the existence of the additional solutions could potentially affect the
bounds. In particular, if mass splittings are significantly different in the new
solutions, the acceptance of the cuts would change and so could the
corresponding exclusion region. The searches are based on hard jets and
missing transverse momentum, and contain lepton vetoes. Thus, if branching
ratios of sparticles into leptons were to change significantly between the
different solutions, then the predicted SUSY signal event rates would also
change, affecting the exclusion regions. 

More recent searches by ATLAS, interpreted as exclusions on the CMSSM
parameter space have used different parameter planes: $A_0=-2m_0$,    $\tan
\beta=10$ or $30$ and $\mu>0$~\cite{ATLAS-CONF-2013-047}. We have checked
these planes for multiple 
solutions on a 101 by 101 grid, not finding any. 
Thus, as far as we can tell, exclusion limits derived upon those planes are
not compromised by additional solutions. 

\subsection{Comparing algorithms}

\begin{figure}
\begin{center}
\includegraphics[scale=0.5]{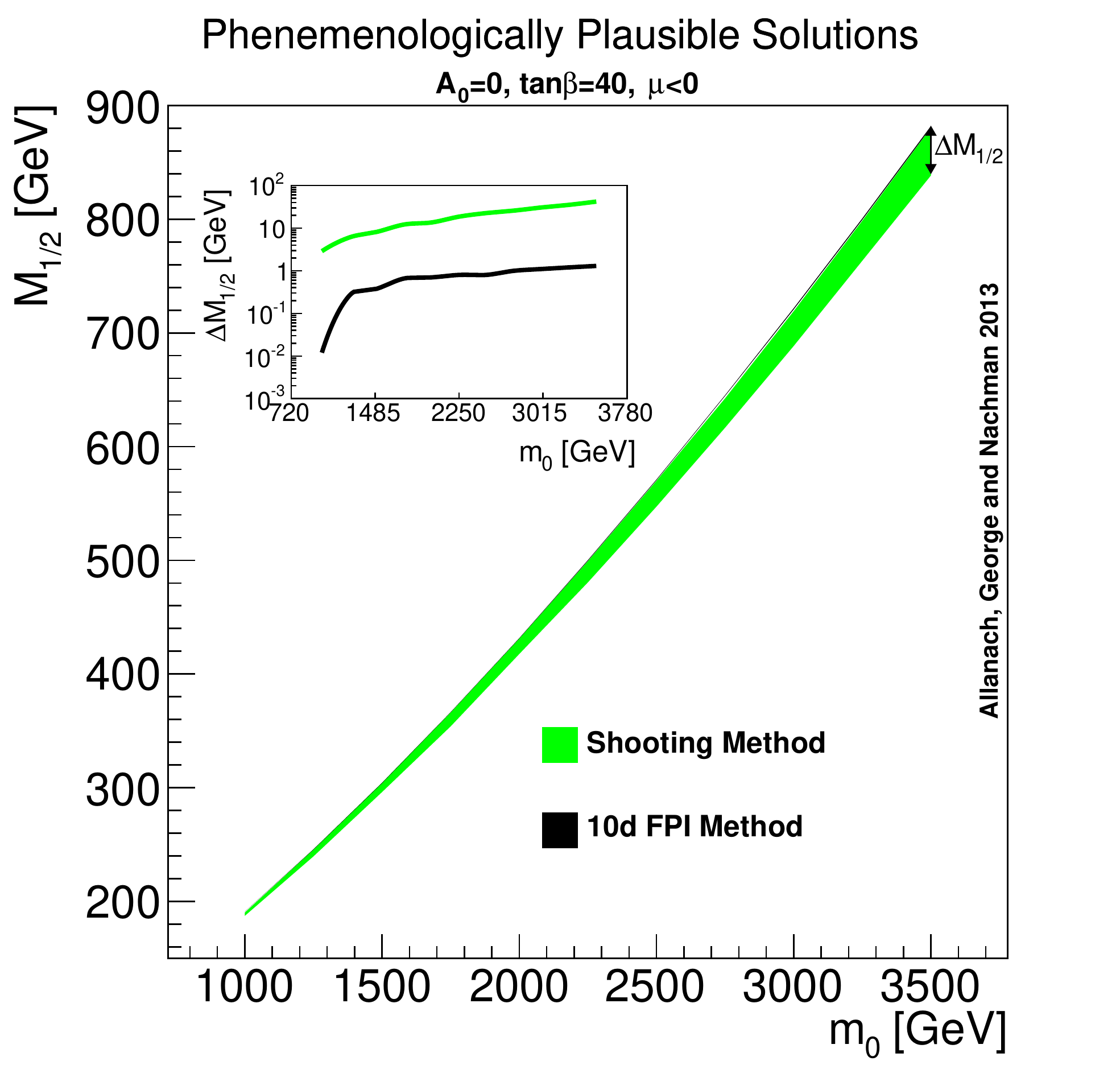}
\end{center}
\caption{A CMSSM extra solution strip for
  $A_0=0,\tan\beta=40$ and $\mu<0$ as computed using a 10d FPI method as in
  Ref.~\cite{Allanach:2013cda} and the 
  shooting method employed in the present paper.
The insert shows the thickness of the strip $\Delta M_{1/2}$ for various
values of $m_0$. \label{fig:thickness}
}
\end{figure}
We now compare our results using the shooting method, to previous estimates
in Ref.~\cite{Allanach:2013cda}. There, $M_Z$ was predicted by initially
relaxing the boundary condition Eq.~\eqref{mucond}, scanning in one
dimension, $\mu(\msusy)$, and using fixed point iteration in the remaining 10
dimensional 
parameter space per CMSSM point. Luckily, the existence of additional
solutions was demonstrated, but in fact many of them could have been missed
due to the fact that FPI was still being used, which could have the concomitant stability
issues  
sketched in Sec.~\ref{sec:fpi}. In fact, the solutions found with the shooting
method are a superset of those found in Ref.~\cite{Allanach:2013cda}.

Comparing the results in Fig.~\ref{fig:numberofsolutions} to equivalent
estimates in Ref.~\cite{Allanach:2013cda} using the different methodology, we see
some differences: we now 
find a much larger area of the parameter plane with multiple solutions. 
For example, we compare maps of a particular strip with two solutions in
Fig.~\ref{fig:thickness}. We shall show below that this strip is of particular
interest because it predicts sparticle masses that are not already excluded by
LEP2.  
The new determination of this multiple solution
region using the multi-dimensional shooting method is shown in green on the
figure, whereas the one-dimensional method of Ref.~\cite{Allanach:2013cda}
only finds the region in the black line. The insert shows that the width of
the strip as measured by $M_{1/2}$ (for a given $m_0$) increases by a factor
of several tens, providing a quantification of how much better the shooting
method performs. 

It is hard to compare the speed of the FPI computation versus that of our
implementation of the shooting method, because often the shooting method will
not find a solution. FPI also does not find additional solutions, and returns
`no-convergence' errors near the boundary of electroweak symmetry breaking. 
On the other hand when the shooting method works, the solution is typically
much more
accurate (i.e.\ $M_Z^\textrm{pred}/M_Z^\textrm{exp}$ is typically much closer
to 1). However, given these caveats, it is still probably helpful to give some
idea of the relative speeds. 
In the case where FPI finds the same solution as our implementation of
the shooting method, FPI is quicker. The vast majority of the run-time is
taken up with running the RGEs between $\mgut$ and $M_Z$. Thus, we may use
this is an approximate unit of time taken for the program to run.
For each iteration of the shooting method we need to run the RGEs from $\mgut$
to $M_Z$ 12 times (one for the central value, and 11 to
determine the 11 forward derivatives in the Jacobian).  This needs to be
repeated typically a few times to converge on a solution (if indeed
it does converge).  
FPI on the other hand, takes around 3-10 iterations, depending upon the
parameter point\footnote{However, points close to the boundary of electroweak
  symmetry breaking can take up to 100 iterations, or indeed never
  converge.}. 
Each of these iterations consists of running from $\mgut$ to $M_Z$ and then
back again, i.e.\ two time units. These rough estimates tell us that the
shooting method is expected to be a few units slower than the FPI method when it
achieves a solution. This is often increased by a factor of 2-10 because 
the shooting method takes that number of shots before a viable solution is
found.

To get a quantitative understanding of the difference in speed of the
algorithms, we performed rectangular scans over part of
Fig.~\ref{fig:charginomass10pos}a for 
three algorithms: 1) pure FPI, 2) FPI to obtain a starting point, perturbing randomly
around this point, then shooting with Newton-Raphson and a forward derivative, and
3) same as 2) but with a symmetric derivative (twice as many evaluations of the RGEs).
We have $\tan\beta=10$, $A_0=0$ and look for positive $\mu$ solutions.  Scanning
with a $31\times31$ grid over the range $m_0=1500-3000$ GeV and
$M_{1/2}=500-800$ GeV algorithms 1, 2 and 3 took 126, 437 and 701 seconds
respectively.  Compared with pure FPI, the shooting method will always take longer
because it does all the computations of FPI (to find the initial guess), plus
additional computation for Newton-Raphson convergence.  It is thus best to summarise
the running times as: the shooting method with forward derivatives added 247\% to
the running time of normal FPI, and with symmetric derivatives, 456\%.  The latter
number is not quite twice the former (as na\"ively expected) since, for example, the
Newton-Raphson matrix inversion does not have to be repeated twice when going from
forward to symmetric derivatives.

Because convergence properties differ over the CMSSM parameter space, we also scanned
a rectangular grid that included part of the no-EWSB region.  We ran a $31\times21$
grid over the range $m_0=3000-4500$ GeV and $M_{1/2}=400-600$ GeV
(with all other parameters the same as in the previous scan).  Per point, this region
ran slower than the previous region.  Algorithms 1, 2 and 3 took 230, 603 and 946
seconds respectively.  Compared with pure FPI, these numbers correspond to an
additional 162\% and 311\% for the forward and symmetric derivative shooting methods,
respectively.

\subsection{The LEP2 limit}

\begin{figure}
\begin{center}
\includegraphics[width=0.7\textwidth]{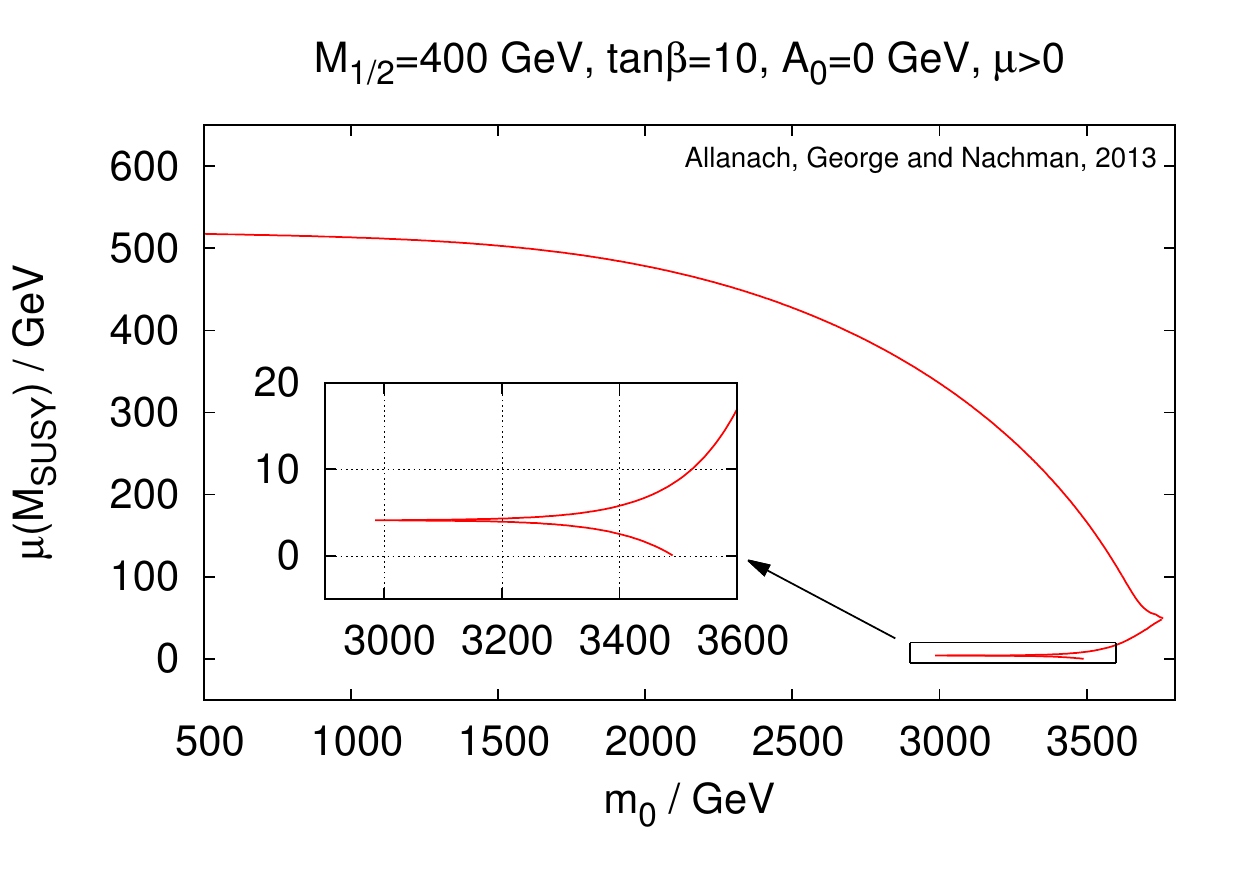}
\end{center}
\caption{A scan over $m_0$ in the CMSSM, solving all 11 low-scale constraints,
  and plotting 
   the value of $\mu(\msusy)$ for all the solutions.  It is clear that the
   values for $\mu(\msusy)$ can differ greatly among the multiple
   solutions. For values of $m_0$ larger than those shown, electroweak
   symmetry is not successfully broken.}
\label{fig:muvsm0}
\end{figure}

We shall see that a sparticle mass bound derived at LEP2 in the context of the
CMSSM imposes strong constraints upon the additional solutions. 
The 95$\%$ CL
lower limit on the chargino mass from LEP2 is 102.5 GeV~\cite{LEP:2001}
within CMSSM-like models. 

The Lagrangian contains the chargino mass matrix $-{\tilde\psi^-}{}^T{\cal
  M}_{\tilde\psi^+}\tilde\psi^+ + h.c.$, 
where~$\tilde\psi^+ = (-i\tilde w^+,\ \tilde h_2^+)^T,\ \tilde\psi^-=
(-i\tilde w^-,\ \tilde h_1^-)^T$, 
${\tilde w}^\pm = ({\tilde w^1} \mp i{\tilde w^2} ) / \sqrt{2}$ for the charged
winos and ${\tilde h_1^-}, {\tilde h_2^+}$ for the charged higgsinos. We also
have, at tree level,
\begin{equation}
{\cal M}_{\tilde\psi^+}\ =\ \left( \begin{array}{cc} M_2 &
\sqrt2\,M_Ws_\beta\\\sqrt2\,M_Wc_\beta & \mu\end{array}\right).
\label{mchi}
\end{equation}
The chargino masses correspond to the singular values of ${\cal
  M}_{\tilde\psi^+}$, i.e.\ the positive square roots of the eigenvalues of
${\cal M}_{\tilde\psi^+}^\dag {\cal M}_{\tilde\psi^+}$~\cite{PDG}:
\begin{eqnarray}
M_{\chi_{1,2}^+}^2 &=& \frac{1}{2} \left\{  \mu ^2 + M_2^2 + 2 M_W^2 \mp
\left[ (\mu^2 + M_2^2 + 2 M_W^2)^2  \right. \right. \nonumber \\
 & & \left. \left. -  4 \mu^2 M_2^2 - 4 M_W^4 \sin^2 2 \beta + 8 M_W^2 
   \mu M_2 \sin 2 \beta \right]^{1/2}
\right\}. \label{charg}
\end{eqnarray}
This places a strong constraint on many of our
multiple solutions, whose values of $\mu$ differ (and therefore the lightest
chargino mass computed by Eq.~\eqref{charg} differs).
In Fig.~\ref{fig:muvsm0} we show the behaviour of $\mu(\msusy)$ for all the
solutions that can be found as $m_0$ changes, for $M_{1/2}=400$ GeV, $A_0=0$,
$\tan \beta=10$ and $\mu>0$ (for this plot we used Newton-Raphson).
For high enough $m_0$, electroweak symmetry breaking
no longer works and so we no longer have any solutions to plot (i.e.\ for
$m_0>3.76$ TeV). 
The appearance of additional solutions near the electroweak symmetry breaking
boundary was qualitatively explained in Ref.~\cite{Allanach:2013cda}. To
summarise, the RGE corrections which change $\mu$ have a larger than usual
effect in this region because $\mu$ approaches zero at the boundary. 
In the figure, we see how $\mu(\msusy)$, derived from minimising the
Higgs potential, starts at a high value at moderate $m_0$, and decreases as
$m_0$ 
is increased and as the boundary of successful symmetry breaking is
approached. This is 
fairly generic behaviour, and applies for different values of $M_{1/2}$. 
For $m_0 \in [3 \textrm{~TeV},\ 3.5 \textrm{~TeV}]$, we obtain two additional
solutions. Between the different solutions, we observe very different values
of $\mu(\msusy)$. The additional solutions have rather low values (again,
generic behaviour near the boundary of electroweak symmetry breaking),
predicting a low $M_{\chi^\pm_1}$ once loop corrections have been added to
Eq.~\eqref{charg}.

\begin{figure}
\begin{center}
\unitlength=\textwidth
\begin{picture}(1,0.5)
\put(0.,0){\includegraphics[width=0.5\textwidth]{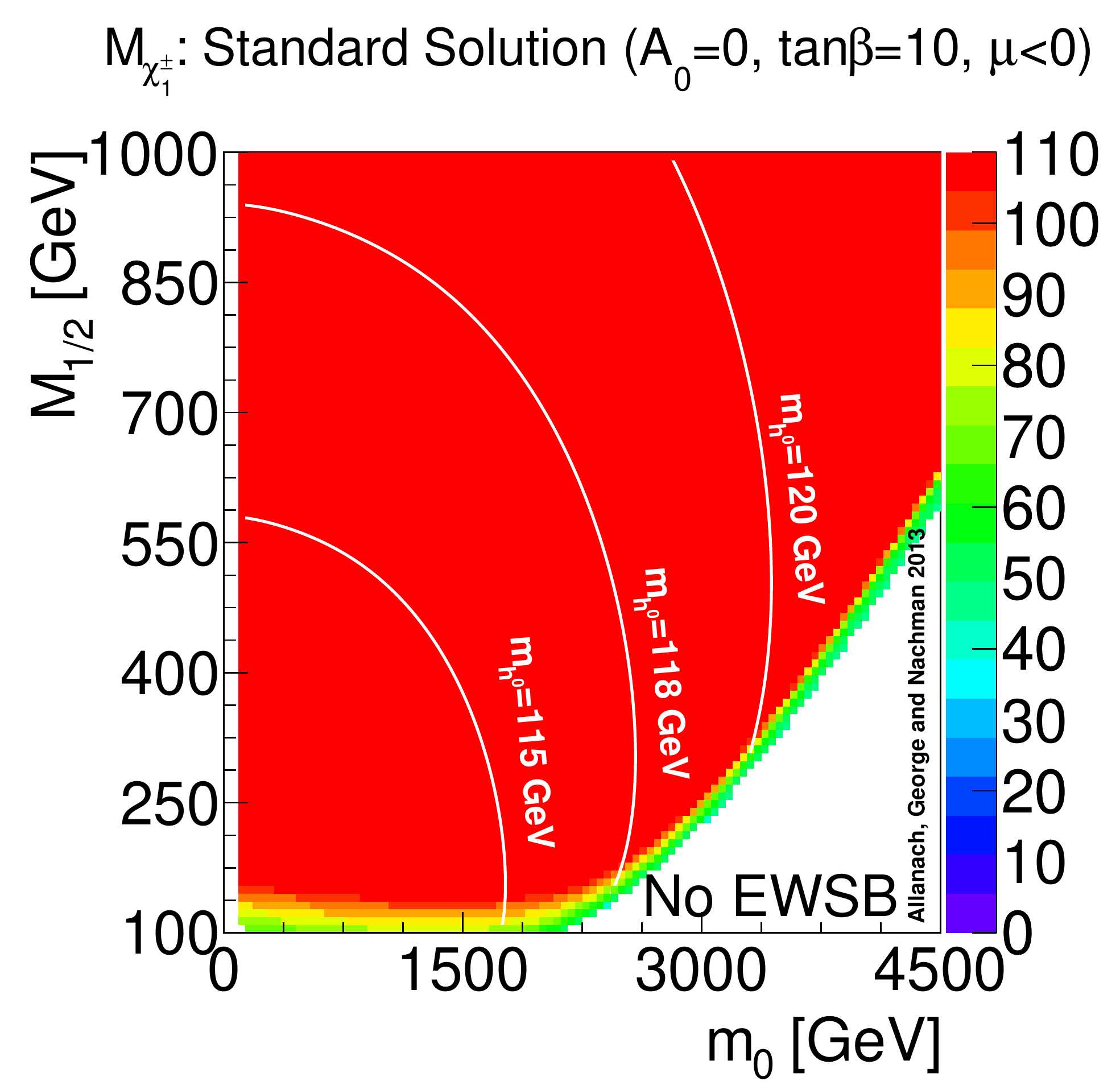}}
\put(0.5,0){\includegraphics[width=0.5\textwidth]{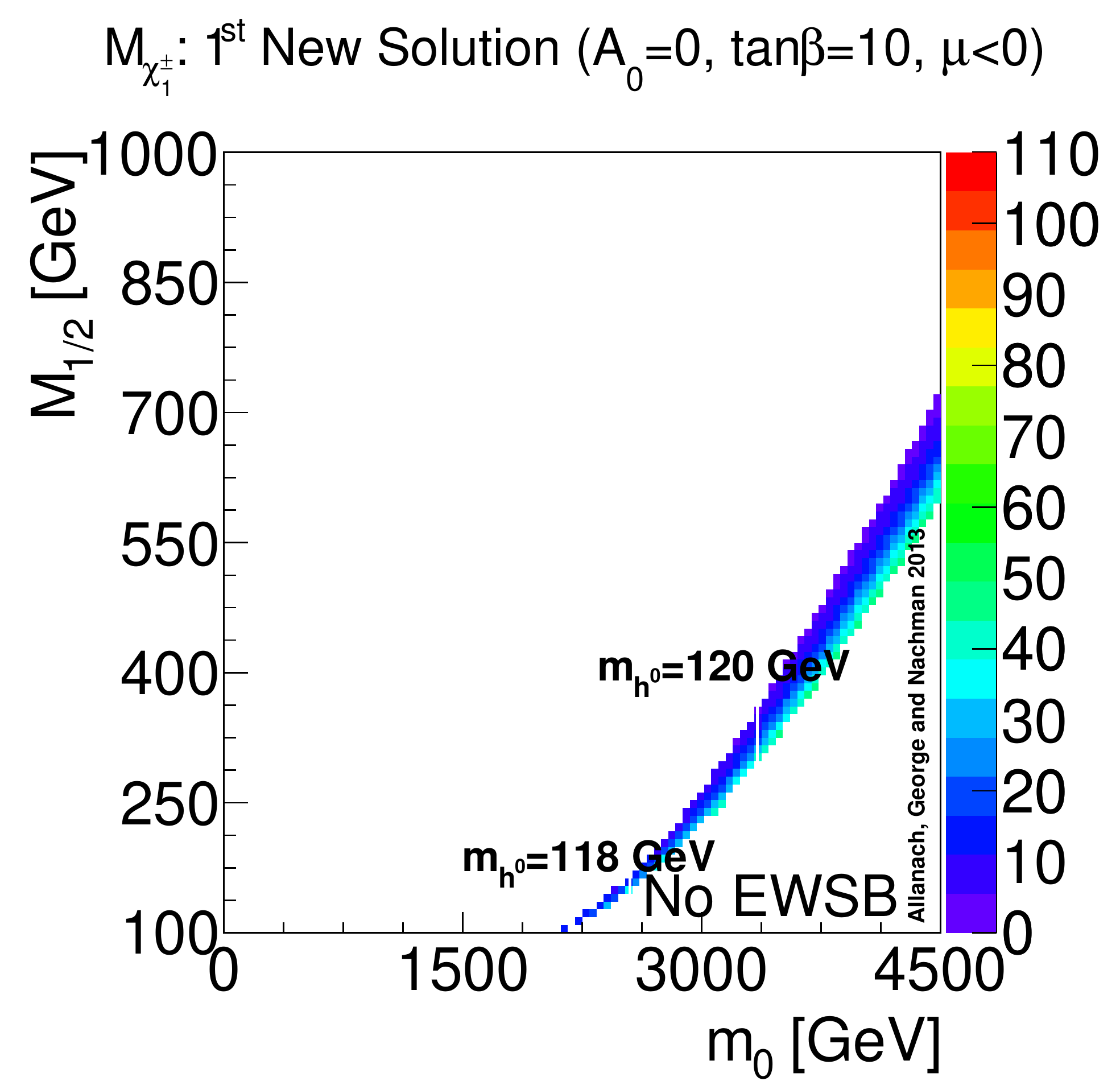}}
\put(0.12,0.38){(a)}
\put(0.62,0.38){(b)}
\end{picture}
\end{center}
\caption{Lightest chargino mass in the CMSSM for $A_0=0$, $\tan\beta=10$
  and $\mu<0$.  White contours are iso-contours of
  lightest CP-even Higgs mass $m_{h^0}$.  The additional solutions in
  (b) are ruled out by the LEP2 lower limit on the
  chargino mass of $102.5$ GeV.
}
\label{fig:charginomass10neg}
\end{figure}

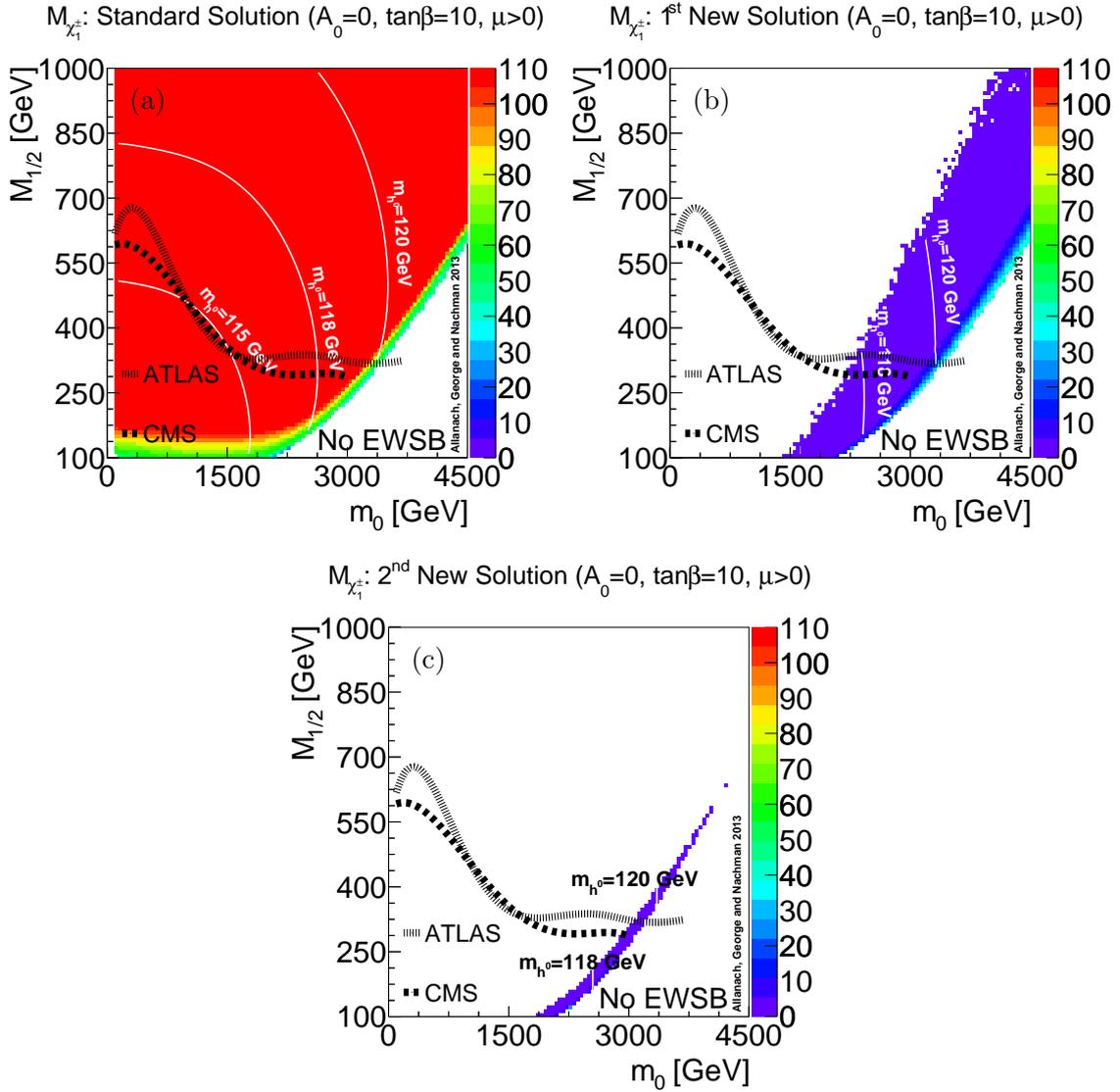
\begin{figure}
\begin{center}
\unitlength=\textwidth
\begin{picture}(1,1)
\put(0,0.5){\includegraphics[width=0.5\textwidth]{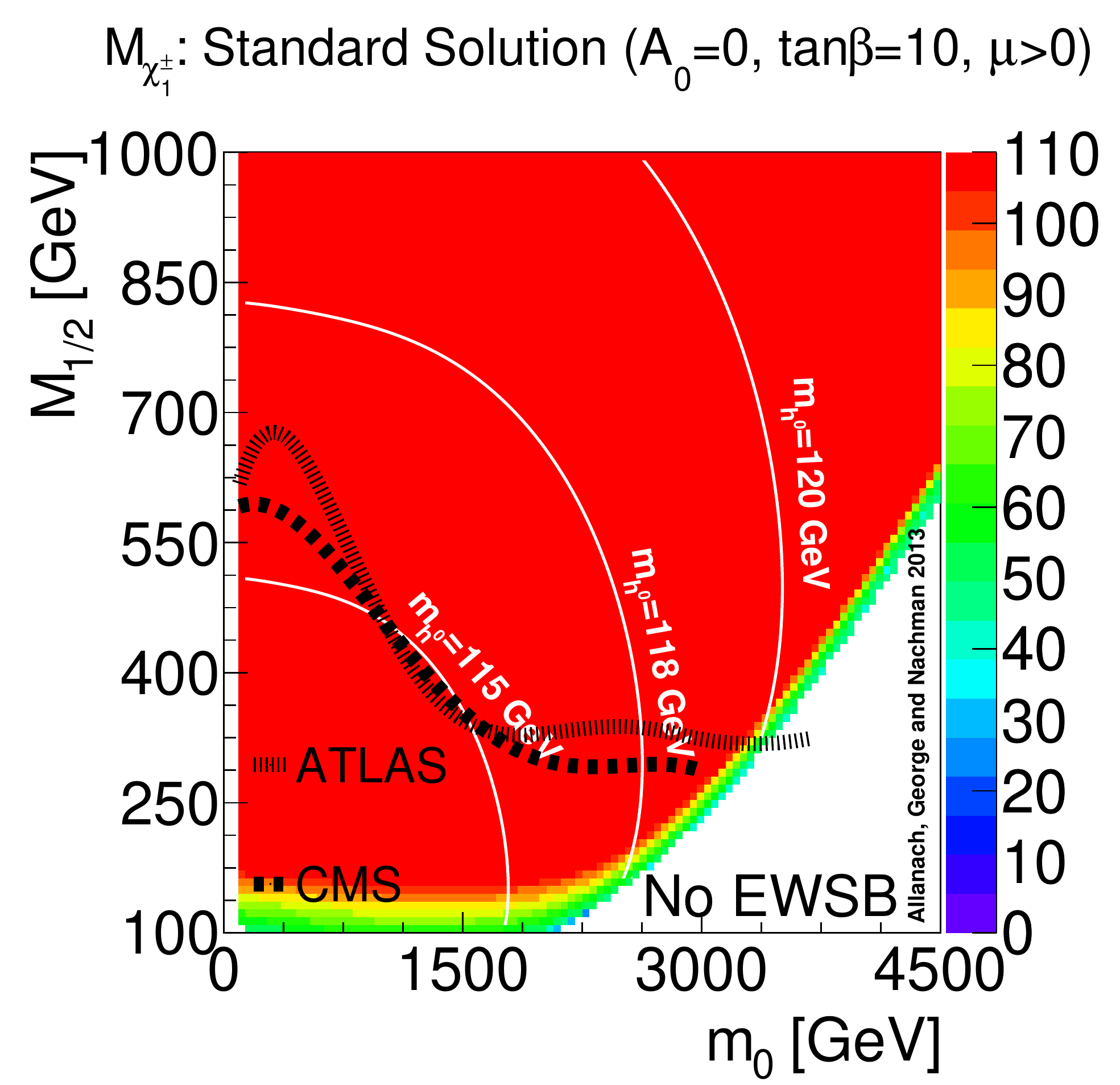}}
\put(0.5,0.5){\includegraphics[width=0.5\textwidth]{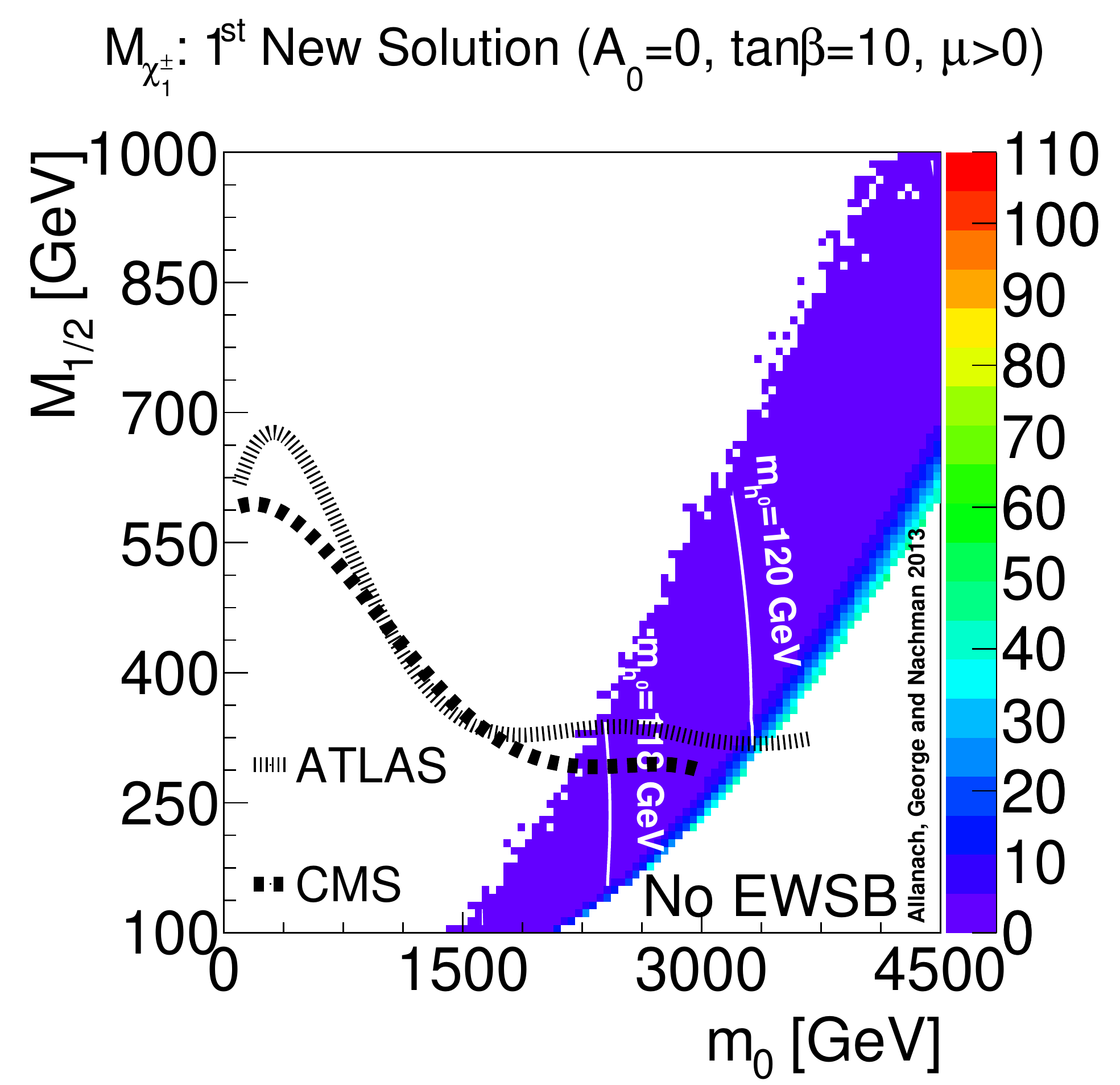}}
\put(0.25,0){\includegraphics[width=0.5\textwidth]{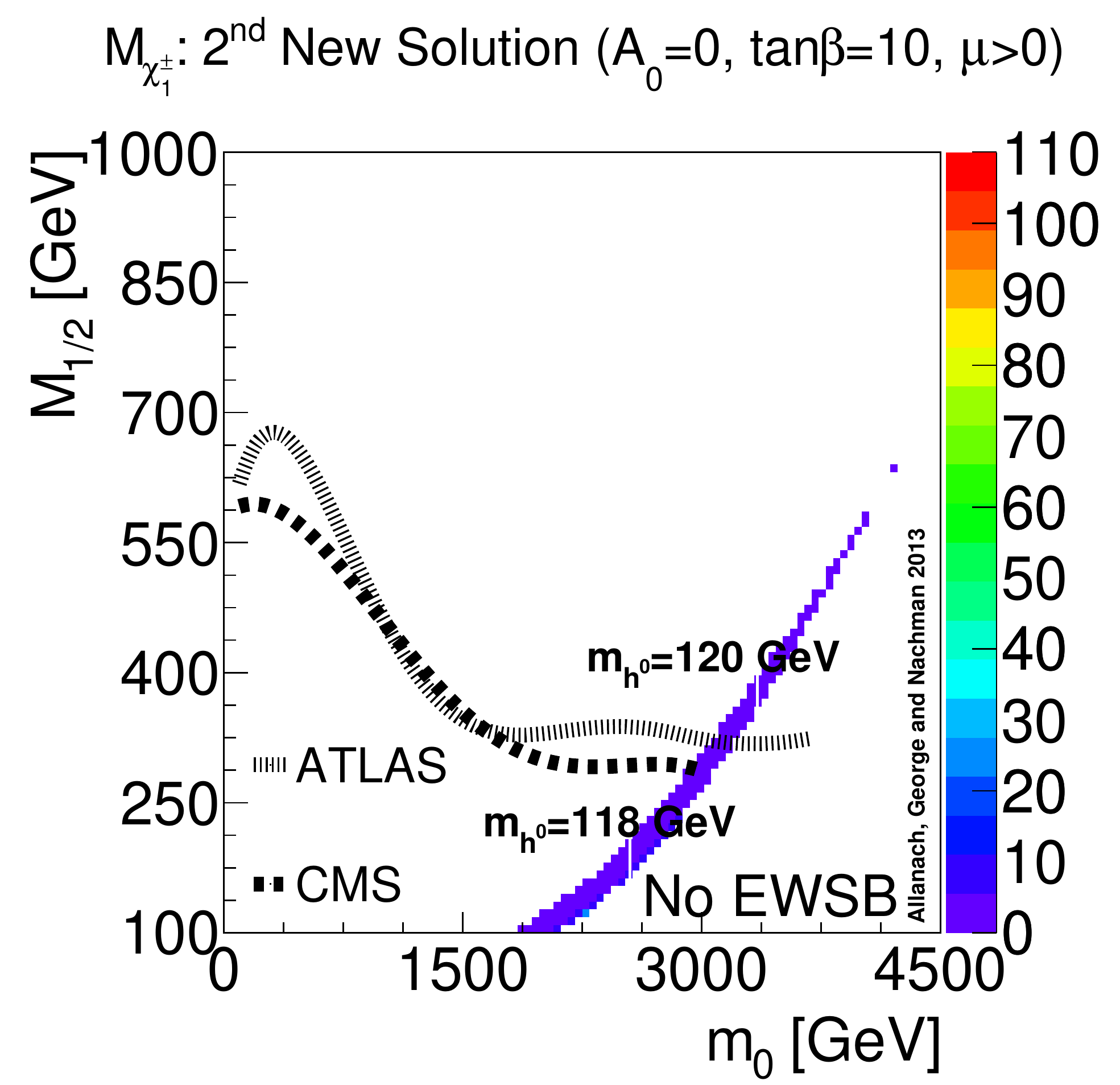}}
\put(0.12,0.88){(a)}
\put(0.62,0.88){(b)}
\put(0.37,0.38){(c)}
\end{picture}
\end{center}
\caption{Lightest chargino mass in the CMSSM for $A_0=0$, $\tan\beta=10$
  and $\mu>0$.  White contours are iso-contours of
  lightest CP-even Higgs mass $m_{h^0}$.  The additional solutions in
  (b) and (c) are all ruled out by the LEP2 lower limit on the
  chargino mass of $102.5$ GeV.
}
\label{fig:charginomass10pos}
\end{figure}

\begin{figure}
\begin{center}
\unitlength=\textwidth
\begin{picture}(1,0.5)
\put(0.,0){\includegraphics[width=0.5\textwidth]{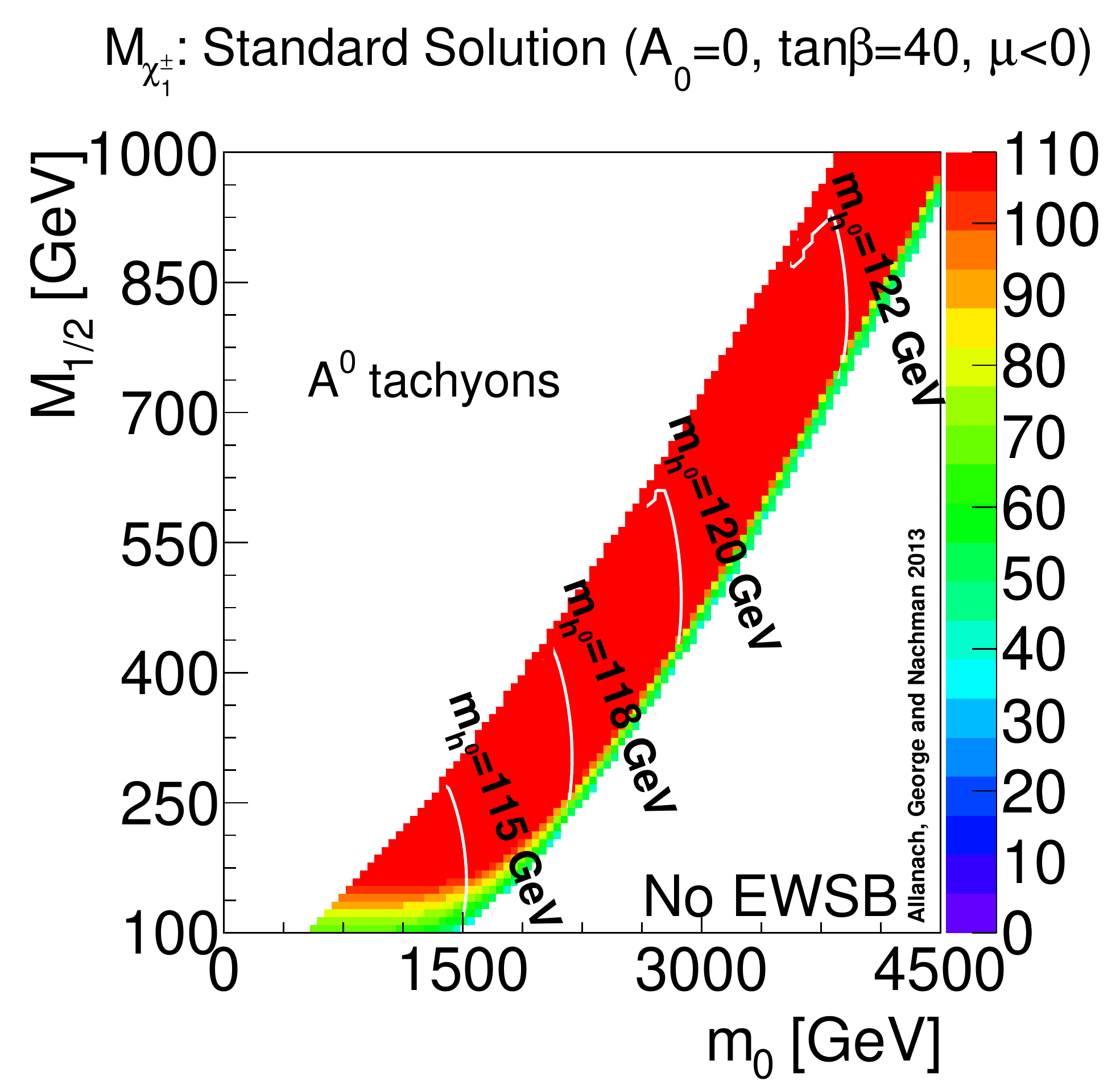}}
\put(0.5,0){\includegraphics[width=0.5\textwidth]{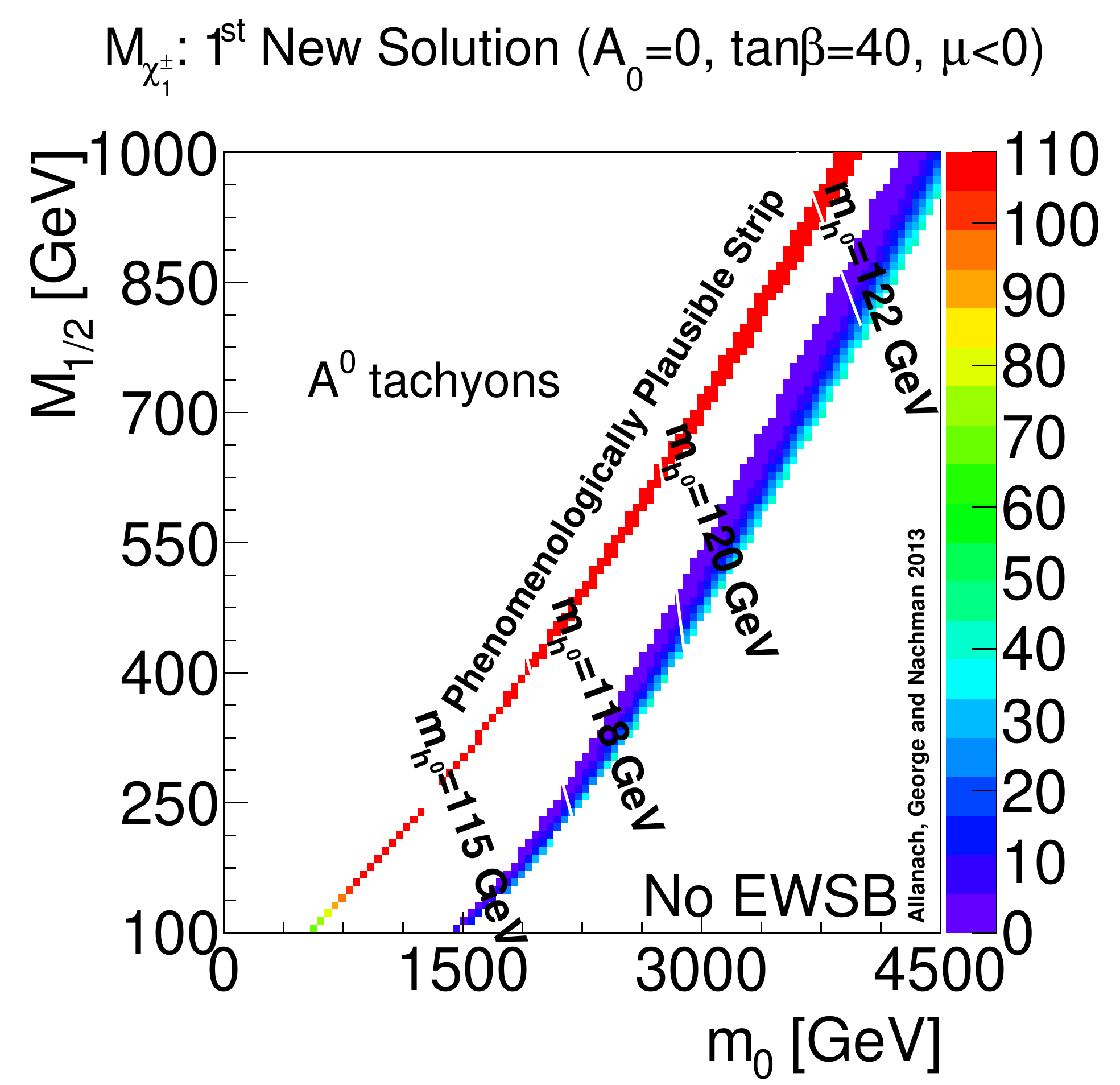}}
\put(0.12,0.38){(a)}
\put(0.62,0.38){(b)}
\end{picture}
\end{center}
\caption{Lightest Chargino mass in the CMSSM for $A_0=0$, $\tan\beta=40$
  and $\mu<0$. White contours are iso-contours of
  lightest CP-even Higgs mass $m_{h^0}$.  (b) has a strip of
  phenomenologically plausible solitions.
}
\label{fig:charginomass40neg}
\end{figure}

\begin{figure}
\begin{center}
\unitlength=\textwidth
\begin{picture}(1,1)
\put(0,0.5){\includegraphics[width=0.5\textwidth]{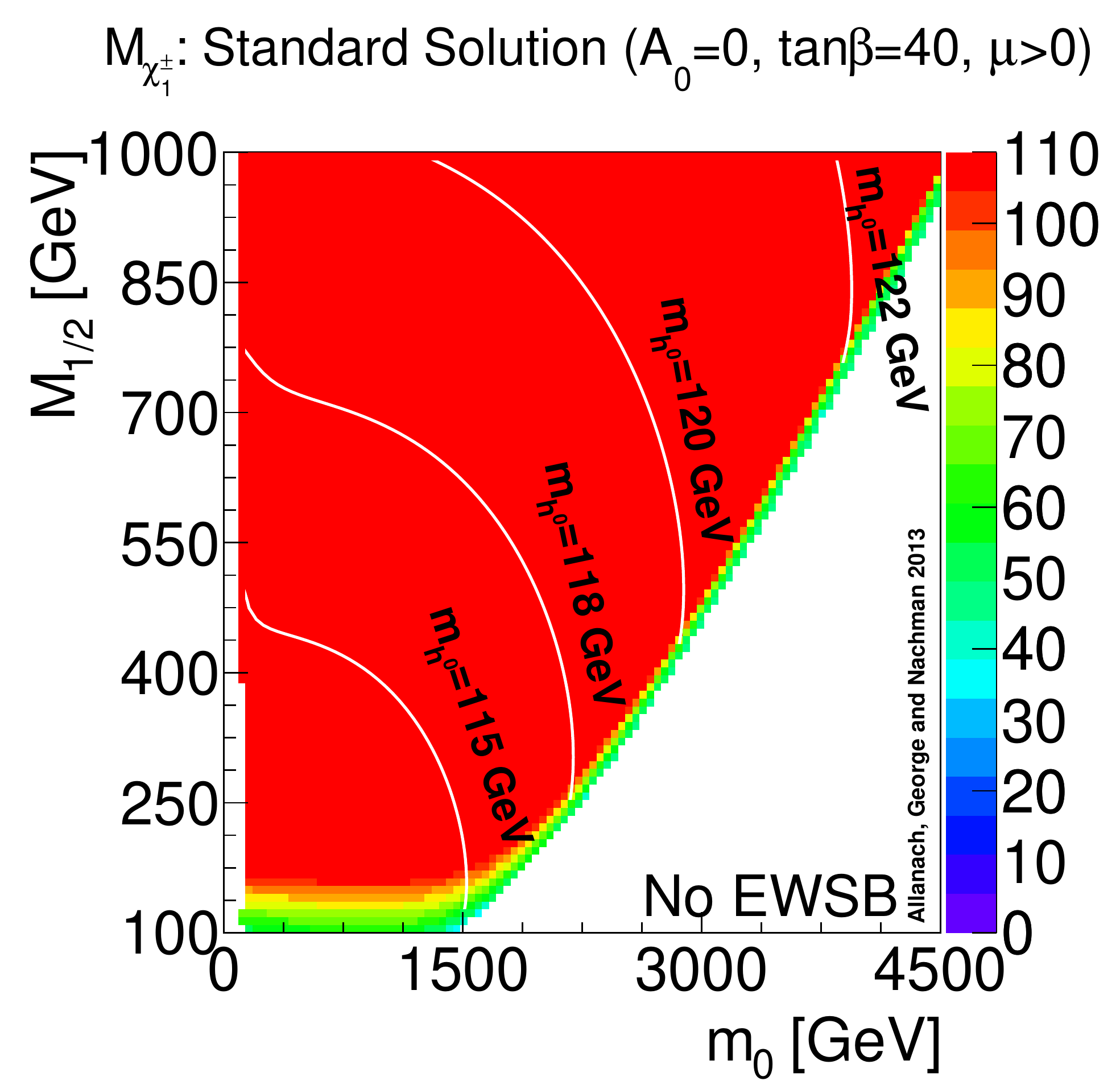}}
\put(0.5,0.5){\includegraphics[width=0.5\textwidth]{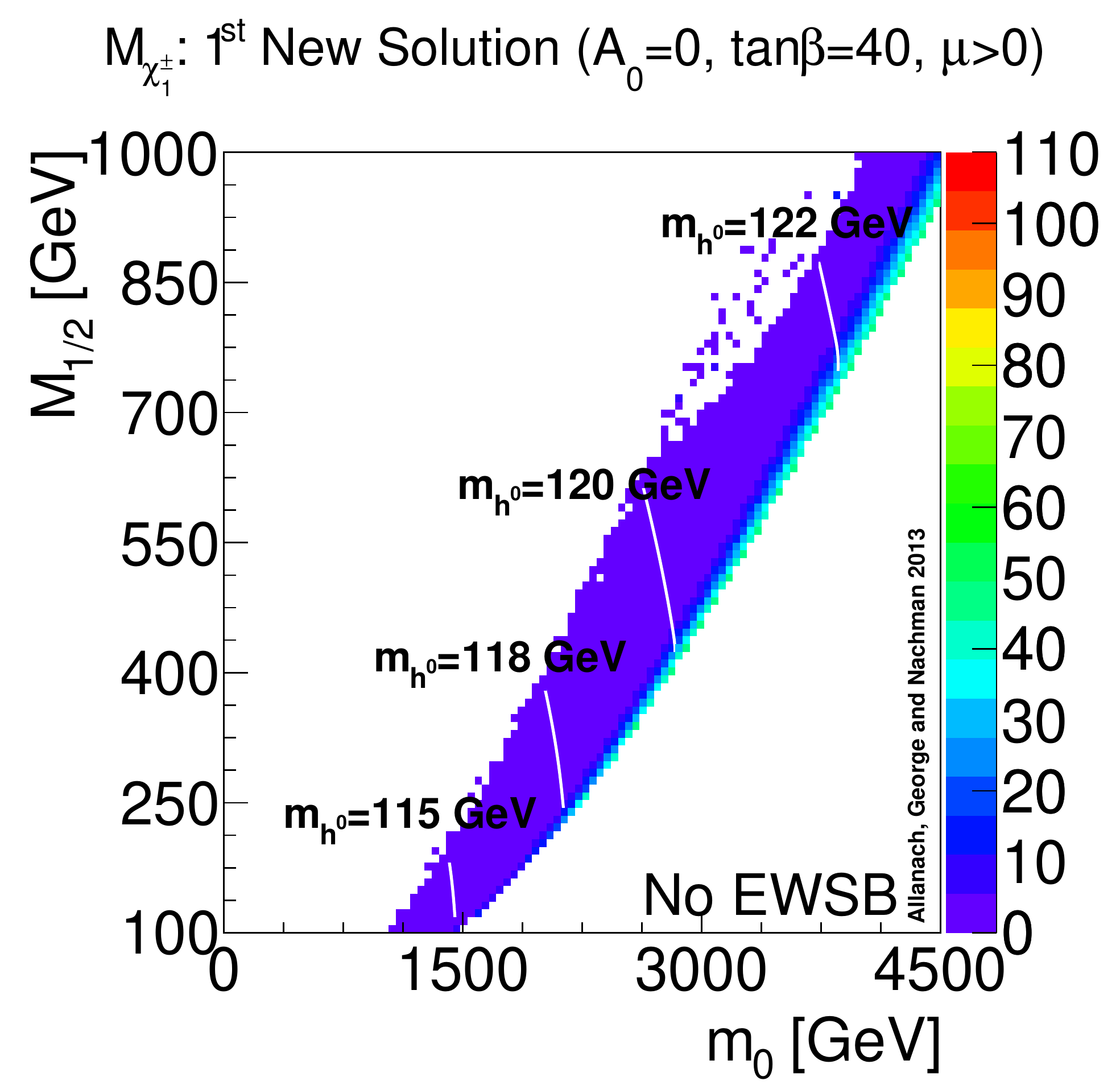}}
\put(0.25,0){\includegraphics[width=0.5\textwidth]{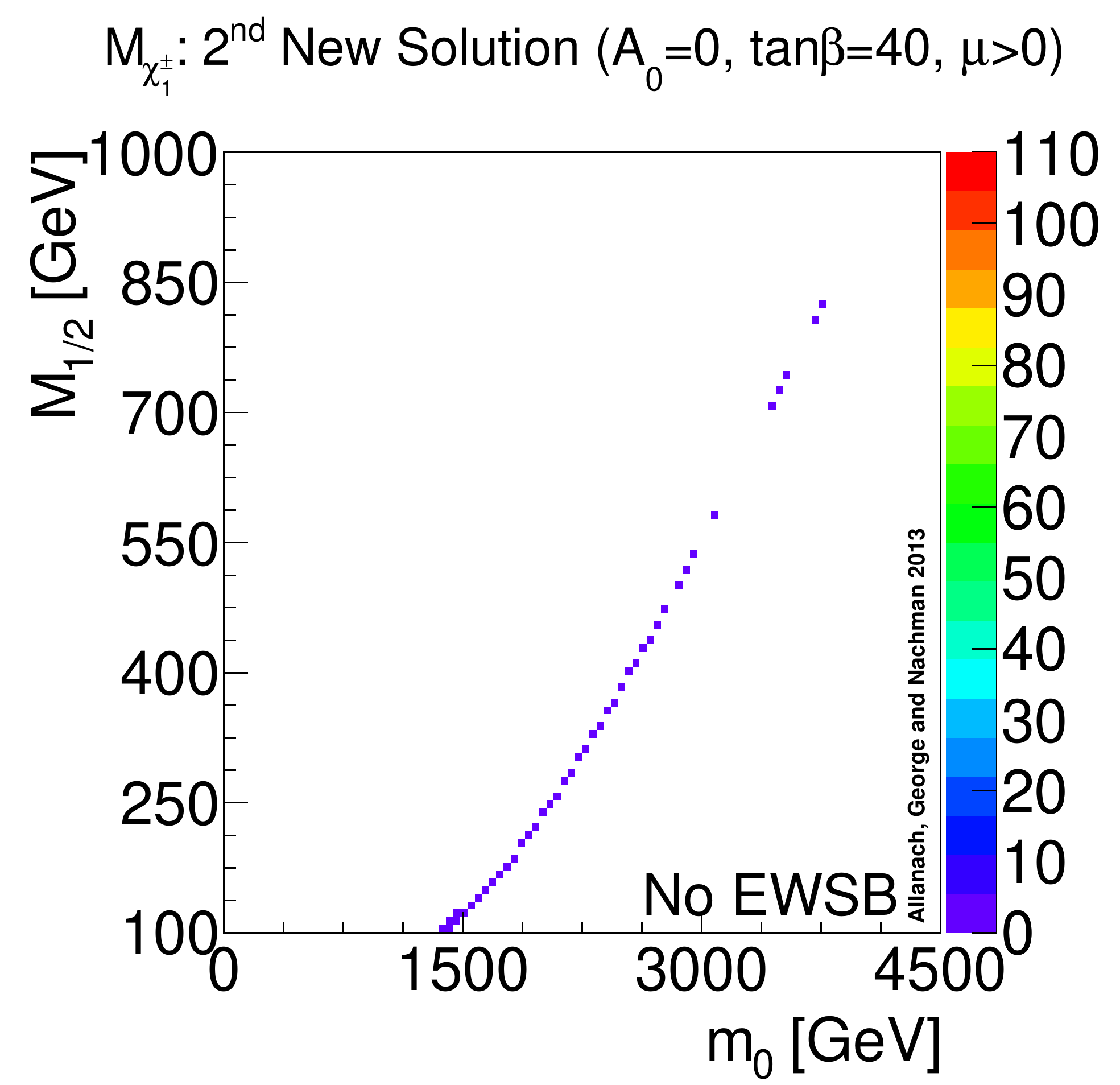}}
\put(0.12,0.88){(a)}
\put(0.62,0.88){(b)}
\put(0.37,0.38){(c)}
\end{picture}
\end{center}
\caption{Lightest Chargino mass in the CMSSM for $A_0=0$, $\tan\beta=40$
  and $\mu>0$. White contours are iso-contours of
  lightest CP-even Higgs mass $m_{h^0}$.  The additional solutions in
  (b) and (c) are all ruled out by LEP2.
}
\label{fig:charginomass40pos}
\end{figure}
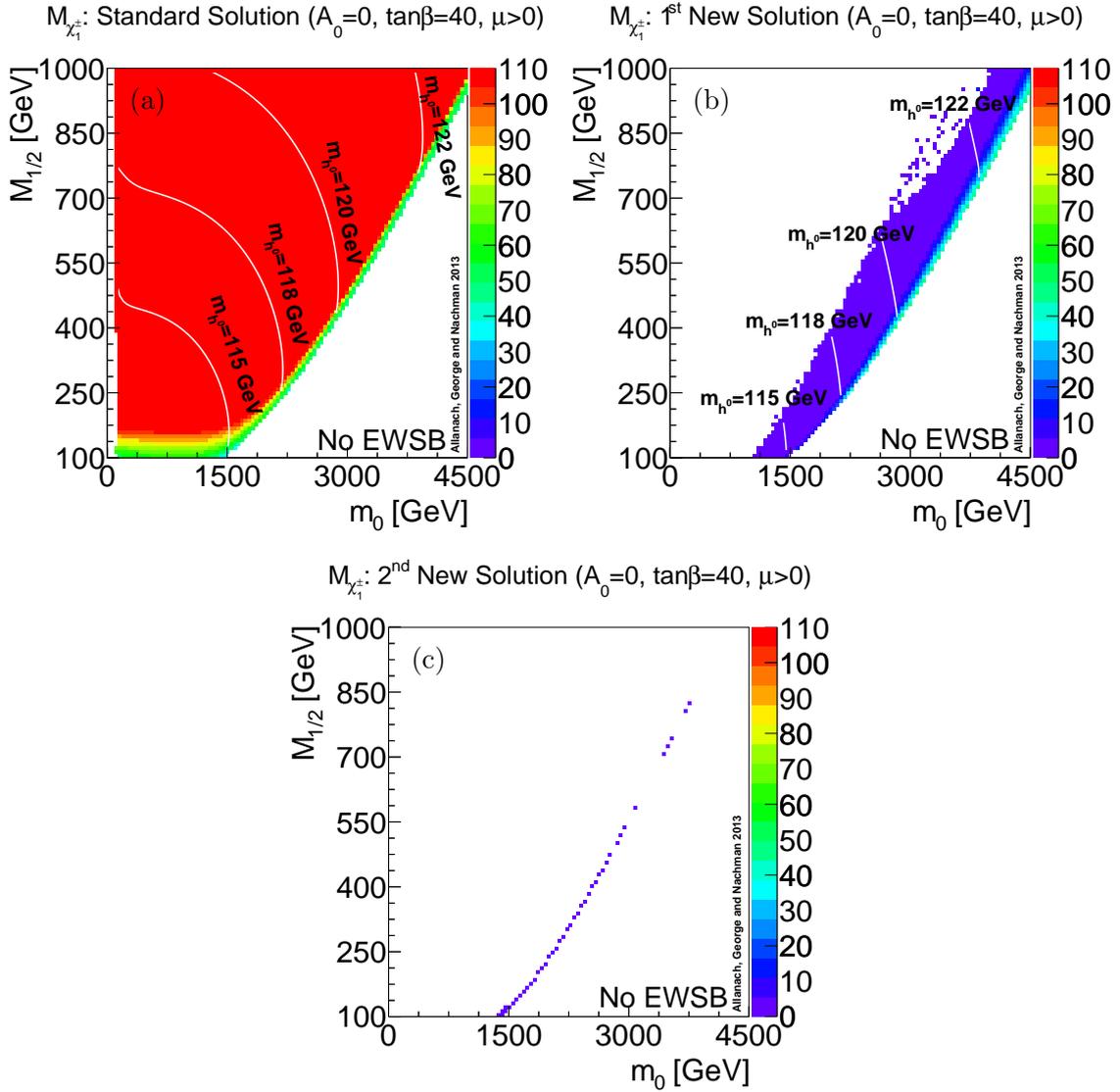

In particular, the multiple solutions near the boundary of
electroweak symmetry breaking have a smaller value of $\mu(\msusy)$ than the
standard solution, with a consequently smaller chargino mass.  
We plot the chargino mass as the background colour of
Figs.~\ref{fig:charginomass10neg}, \ref{fig:charginomass10pos}, \ref{fig:charginomass40neg}
and~\ref{fig:charginomass40pos}
for the standard solutions and the new solutions 
found by our shooting algorithm using Broyden's method. 
We see from Figs.~\ref{fig:charginomass10neg}a,~\ref{fig:charginomass10pos}a
that, for the standard solutions, only a
small region right next to the EWSB boundary has $M_{\chi_1^\pm}<102.5$ GeV,
and falls afoul of the LEP2 chargino mass constraint for $\tan \beta=10$ and
either sign of $\mu$.  
Figs.~\ref{fig:charginomass10neg}b, \ref{fig:charginomass10pos}b and~\ref{fig:charginomass10pos}c
show that the multiple solutions all {\em 
  fail}\/ the  LEP2 chargino mass constraint for $\tan \beta=10$ and
$A_0=0$. Once this is taken into account, the additional solutions do not
resurrect any points that were already ruled out by the 95$\%$ CL region of
ATLAS and CMS, as shown in Fig.~\ref{fig:charginomass10pos}, for example.  
The LHC limits are therefore unchanged on  the $\tan\beta=10,\mu>0$ plane.  

More
generally, from
Figs.~\ref{fig:charginomass10neg}, \ref{fig:charginomass10pos}, \ref{fig:charginomass40neg}
and~\ref{fig:charginomass40pos}
we see that 
every additional solution in the class adjacent to the no EWSB region {\em
  fails}\/
the LEP2 chargino mass constraint for $A_0=0$, $\tan \beta=10$ and $\tan
\beta=40$ and 
either sign of $\mu$. This means that the only viable additional
solution strip shown in these plots is the one at higher values of $M_{1/2}$ in
Fig.~\ref{fig:charginomass40neg}b. We refer to this as the ``phenomenologically
plausible strip''.

We also show iso-contours of the lightest CP even Higgs mass in
Figs.~\ref{fig:charginomass10neg}, \ref{fig:charginomass10pos}, \ref{fig:charginomass40neg}
and~\ref{fig:charginomass40pos}. 
One should bear in mind an estimated 3 GeV error upon the prediction coming
from higher order corrections. Much of the available parameter space is ruled
out by the recent LHC measurements of a boson consistent with a 125 GeV
Higgs~\cite{Aad:2012tfa,Chatrchyan:2012ufa}. 
While this bound is not the main focus of the present paper, we
see from Fig.~\ref{fig:charginomass40neg}b that there is an allowed region with
multiple solutions: the
phenomenologically plausible strip
with $m_{h^0}>122$ GeV. 

\subsection{The phenomenologically plausible strip}

The only region of parameter space investigated that has additional solutions
which are not excluded outright
by the LEP2 chargino bounds is the thin phenomenologically plausible strip
(the upper region in Fig.~\ref{fig:charginomass40neg}b, also shown in
Fig.~\ref{fig:thickness}).  In this region, the $\overline{DR}$ CP-odd neutral
Higgs mass $m_{A^0}$ 
tends toward zero, enhancing its loop correction in threshold effects (for
example to Yukawa and gauge couplings). This enhancement allows for larger
than usual corrections, leading to the presence of multiple solutions.  For
more detail see Ref.~\cite{Allanach:2013cda}. 
We shall now
study the phenomenological properties of solutions along this strip in more
depth. In particular, we wish to know how much the sparticle masses and other
observables differ between the different solutions. If sparticle masses differ
by less than 1$\%$ or so, then the solutions are similar enough such that LHC
exclusion regions are unlikely to be different for the separate solutions. 
However, if there are significant differences in masses leading to different
kinematics of SUSY signal events, cut efficiencies may be different enough to
alter exclusion regions. We should bear in mind, however, that a future linear
collider could be sensitive to fractional
mass differences even at the sub-percent level, particularly in the
slepton/chargino/neutralino sectors~\cite{Allanach:1463554}. Thus,
even sub-percent differences could be relevant for future searches or
measurements. The branching ratios of sparticle decays may also be affected by
changes in masses and MSSM couplings. These may then in turn affect event
rates for the production of particular final states. 
It was already shown in Ref.~\cite{Allanach:2013cda} that the values of
$\mu(\msusy)$ and $m_3^2(\msusy)$ vary more than other parameters between the
different solutions. The masses of MSSM particles that are most sensitive to
$\mu(\msusy)$ and 
$m_3^2(\msusy)$ are: neutralino\footnote{It is the higgsino mass that 
  depends upon $\mu(\msusy)$, and so only those neutralinos that have a
  significant higgsino component will show a significant sensitivity. Near the
  boundary of bad electroweak symmetry breaking at high $m_0$, there is
  sensitivity in the 
  lightest neutralino mass, but for lower $m_0$, only the heavier neutralinos
  show any mass changes.} and
chargino masses, third generation 
sparticle masses and the pseudo-scalar Higgs boson mass. We shall display some
of these differences below. 
We shall also display 
branching ratios of stop and sbottom decays into charginos because they
illustrate some significant differences between the two solutions. 

\begin{figure}
\begin{center}
\includegraphics[scale=0.4]{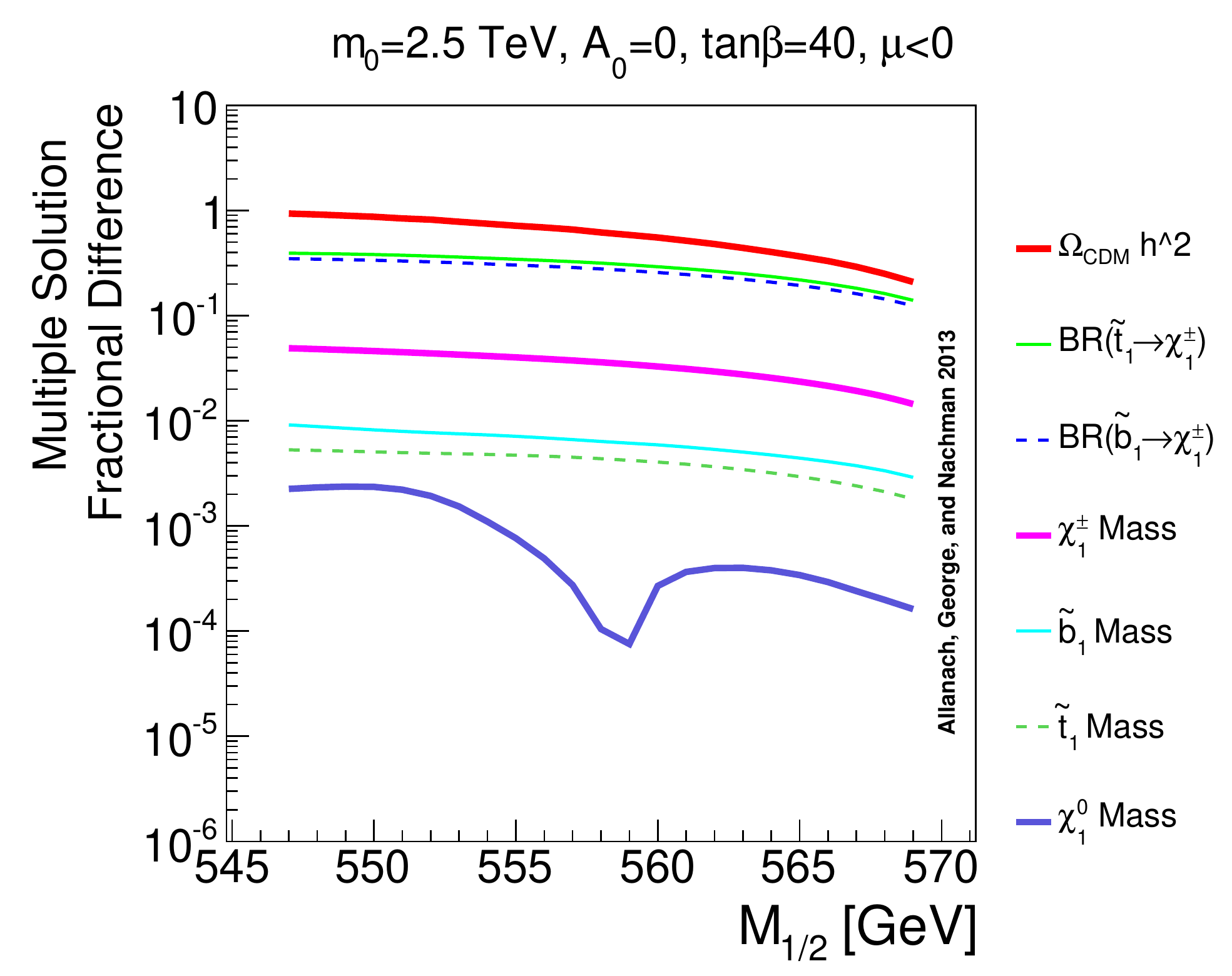}\includegraphics[scale=0.4]{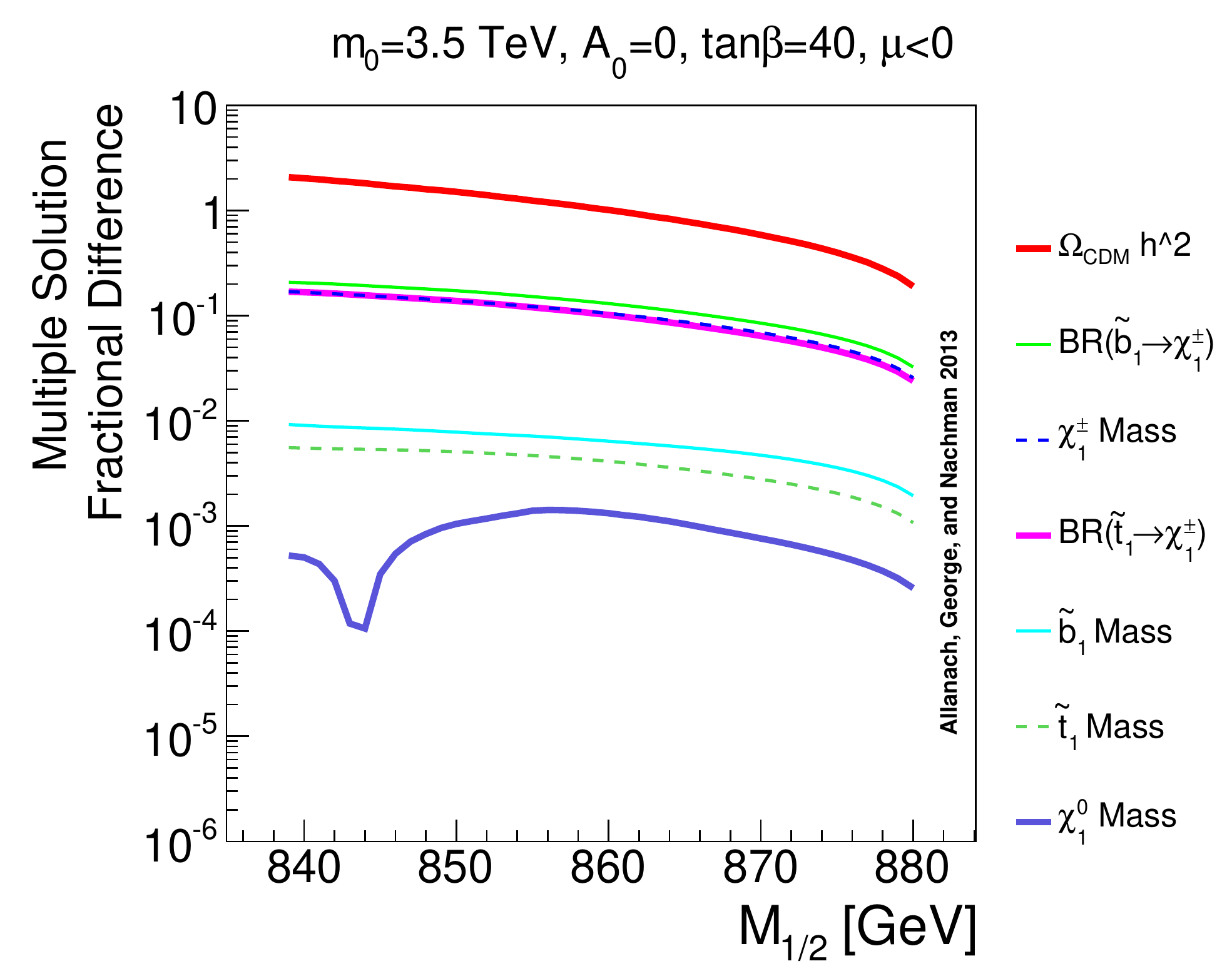}
\end{center}
\caption{Fractional differences between the multiple solutions for various
  phenomenological properties at two slices of the phenomenologically plausible
  strip.  The ordering in the legend matches the ordering of the plotted lines.}
\label{fig:BandWidth}
\end{figure}

Fig.~\ref{fig:BandWidth} shows how  the fractional difference between
the multiple solutions for various phenomenological properties varies across
the strip 
for two fixed values of $m_0$.  The MSSM spectrum and coupling information was
passed with the SUSY Les Houches Accord format~\cite{Skands:2003cj} to
{\tt SUSY-HIT}~\cite{Djouadi:2006bz}, which calculated the branching ratios.
Dark matter properties were then computed
using {\tt micrOMEGAs}~\cite{Belanger:2006is,Belanger:2008sj}.  For a given
calculated phenomenological parameter $X$, the fractional difference is
defined as $|X_1-X_2|/X_1$ where $X_1$ is the predicted value of $X$
corresponding to the solution with a higher chargino mass and $X_2$
corresponds to the solution with the lower chargino mass.  While the there are
only sub-percent level differences between the masses of the lightest
third-generation squarks and the lightest neutralinos, there are significant
deviations for other quantities.  The most significant differences are in
the predicted thermal relic density of neutralino cold dark matter,
$\Omega_\text{CDM} h^2$, in which 
there can be 100\% differences, depending upon the universal model parameters.  
There are also large differences at the tens-of-percent level in the branching
ratio of the third generation squarks into charginos.  At high $m_0$, the
chargino mass itself can also vary at the 10\% level, which would necessarily
affect the limits placed on this region of parameter space for analyses
employing charginos in decays~\cite{Bai:2013ema}.  At lower values of
$m_0$, 
$M_{\chi^\pm_1}$ differs only at the percent level and so we would not expect
there to be a large impact on limits.  Thus, even though the chargino mass is
large enough along the strip for the solutions to be viable, for a 
portion of this strip the phenomenological properties are so close that
excluding one solution would also result in excluding the additional
solution. 

\subsection{Example of Dark Matter Differences}
Now we exemplify the most important difference between the solutions: that
of a different predicted thermal relic density of dark matter in the universe.
Recently, data from the Planck satellite have been used to derive the
constraint~\cite{Ade:2013zuv}
\begin{equation}
\Omega_{CDM} h^2=0.1198 \pm 0.0026.
\end{equation}
We place a dominant
theoretical uncertainty on our prediction of 0.01 coming from
loops (the thermal relic density is only calculated by {\tt micrOMEGAs}~to
tree-level order), and therefore require the predicted 
thermal relic density of neutralinos to be 
$\Omega_{CDM} h^2 \in [0.0998,\ 0.1398]$.
After a brief scan, we found a 
parameter point in the phenomenologically plausible strip ($m_0=760$ GeV, $M_{1/2}=141.72$ GeV, $A_0=0$,
$\tan \beta=40$, $\mu<0$)  
where the standard solution predicts $\Omega_{CDM} h^2=0.34$, i.e.\ well outside of
this range, but 
where the {\em additional} solution prediction of $\Omega_{CDM} h^2=0.118$ is near
the central value. It turns out that this point has the $\chi_1^0$ mass being
approximately half of the Higgs mass. The $\chi_1^0$ mass, which changes
slightly between the solutions, is more exactly half the lightest CP even
Higgs mass for the 
additional solution, which leads to very efficient annihilation of neutralinos
through an $s-$channel $h^0$ into quark or lepton pairs, significantly
reducing the relic density from 0.34 to 0.118.

\subsection{Explorations in parameter space}

It is possible to relax one or more of the constraints $f_i$ at the low scale
and solve only a subset of these equations.  Doing so will highlight the existence
of multiple solutions, and give insight into how much the low-scale predictions
change with respect to input parameters.
For each constraint that is relaxed, there will be one high-scale parameter $V_j$
(it doesn't matter which one) that is no longer determined by the Newton-Raphson
solver.
Such high-scale parameter(s) are controlling parameters, in that we now fix
them for the solution of that point.  To get a feeling for how the parameter
space looks we can scan these parameters~--- fix them at successive values~---
and plot the resulting function $f_i$ which has been left unconstrained.

Here, we choose to relax the $\mzpred^2$ constraint (i.e.\ $f_1$ can take any value)
and the single controlling high-scale parameter is $V_5=[\mu(\mgut)/1000\mbox{~GeV}]^2$.
We scan $V_5$ over a large range and use the Newton-Raphson method to solve for
the remaining 10 parameters using the remaining 10 constraints.  Because the
parameter space is large, we utilise a ``line walking'' technique in order to
improve the efficiency of finding solutions for a given $V_5$.  Given two solutions
that satisfy the 10 low-scale constraints, and are close in $V_5$ at the high-scale,
we can use a linear approximation on the $V_i$ to project and make a guess at the
high-scale parameters for the solutions either side of these original 2.  Doing this
successively allows us to ``walk'' through the high-scale parameter space,
using $V_5$ as the parameter of the line, finding adjacent points that satisfy
the 10 low-scale constraints.  The initial point used is the one found by the
standard FPI method.

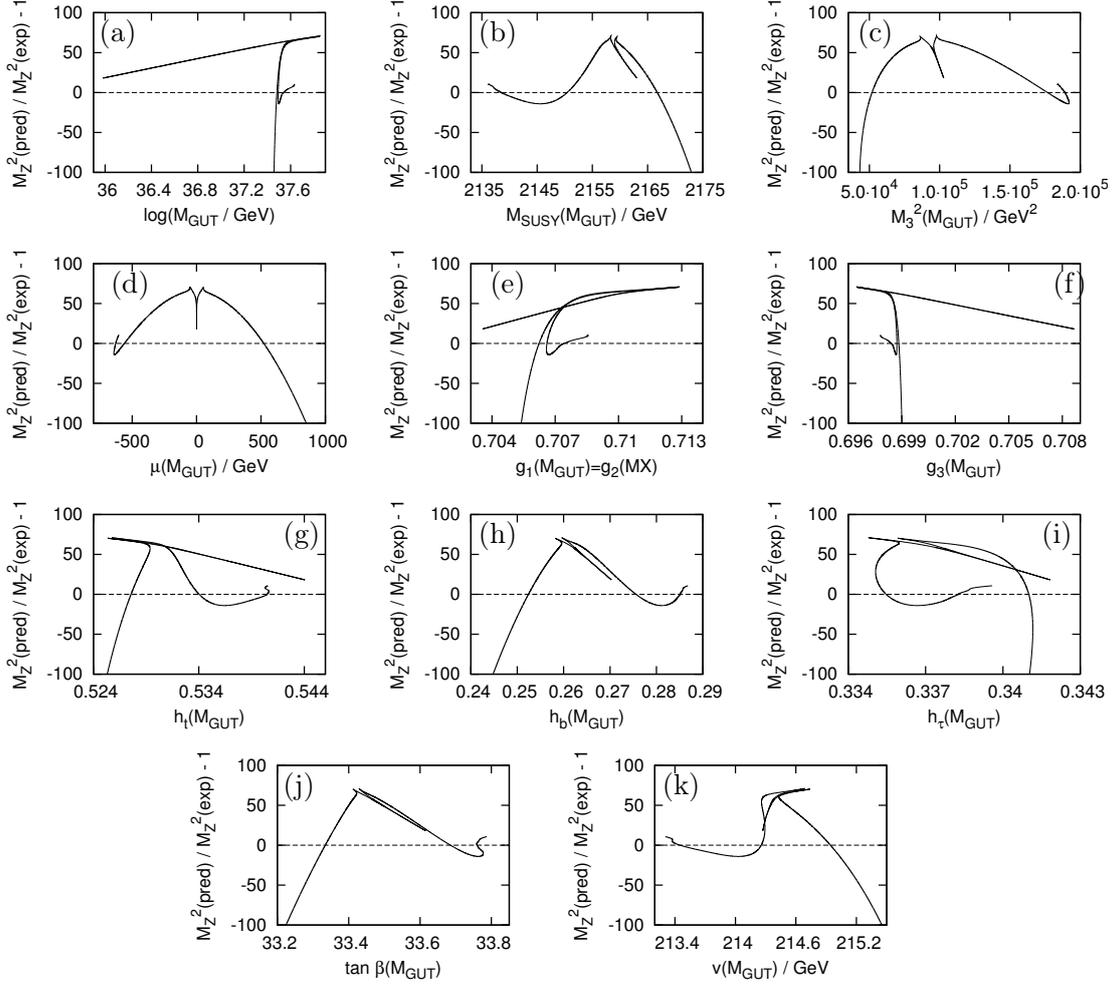
\begin{figure}
\begin{center}
\unitlength=\textwidth
\begin{picture}(1,0.88)(0,0)
\put(0,0.66){\includegraphics[width=0.32\textwidth]{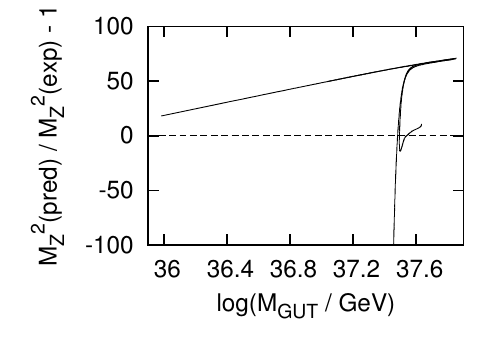}
\includegraphics[width=0.32\textwidth]{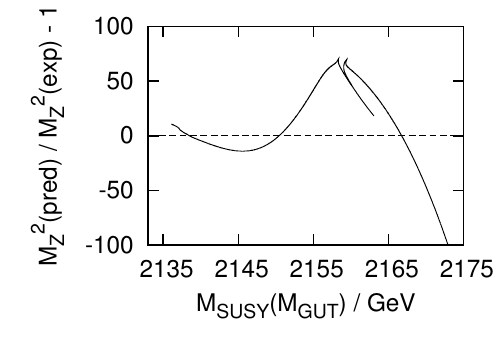}
\includegraphics[width=0.32\textwidth]{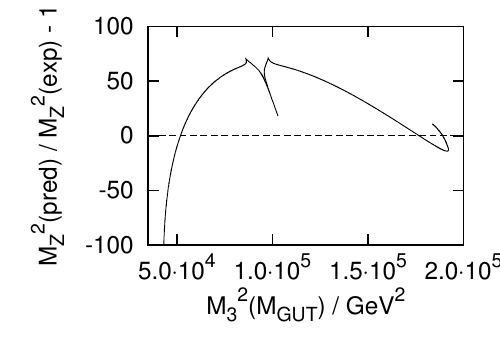}}
\put(0,0.44){\includegraphics[width=0.32\textwidth]{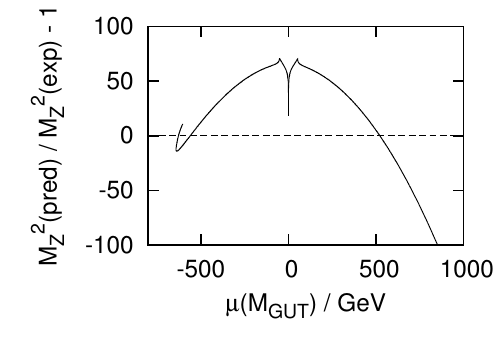}
\includegraphics[width=0.32\textwidth]{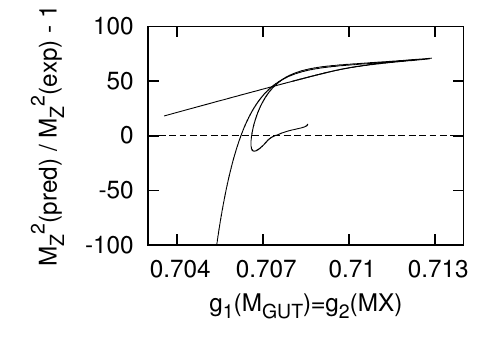}
\includegraphics[width=0.32\textwidth]{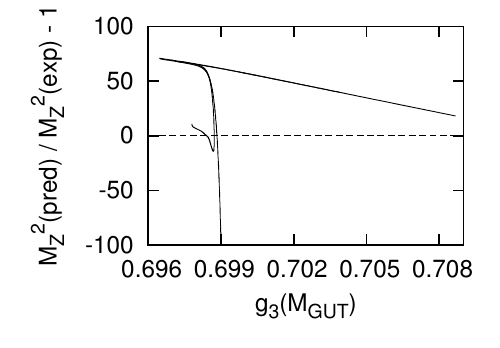}}
\put(0,0.22){\includegraphics[width=0.32\textwidth]{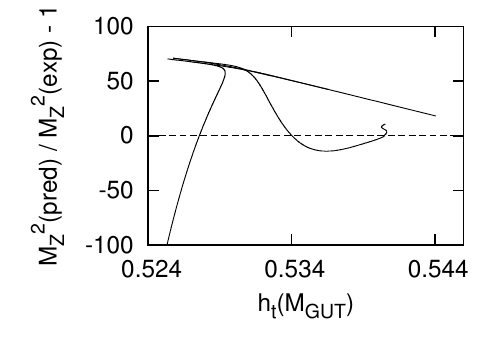}
\includegraphics[width=0.32\textwidth]{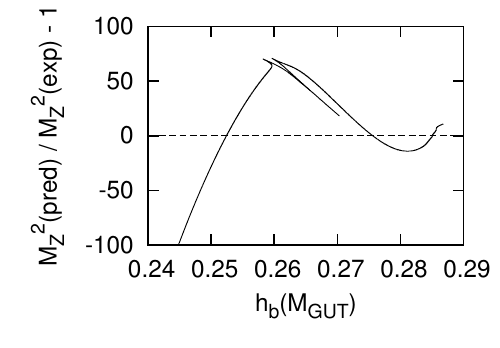}
\includegraphics[width=0.32\textwidth]{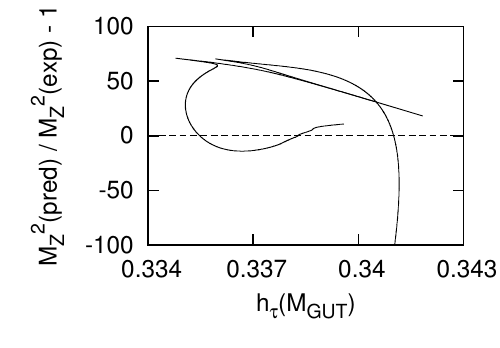}}
\put(0.16,0){\includegraphics[width=0.32\textwidth]{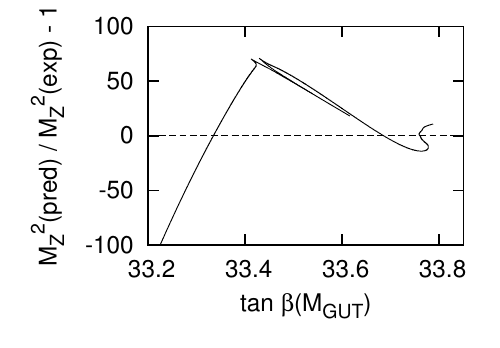}
\includegraphics[width=0.32\textwidth]{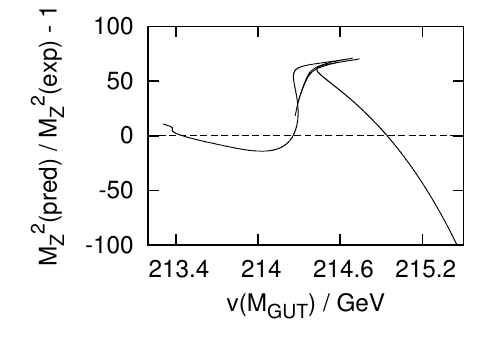}}
\put(0.10,0.84){(a)}
\put(0.43,0.84){(b)}
\put(0.76,0.84){(c)}
\put(0.11,0.62){(d)}
\put(0.44,0.62){(e)}
\put(0.93,0.62){(f)}
\put(0.26,0.4){(g)}
\put(0.43,0.4){(h)}
\put(0.92,0.4){(i)}
\put(0.26,0.18){(j)}
\put(0.59,0.18){(k)}
\end{picture}
\caption{The 11 projections of a function defined on the 11-dimensional
    high-scale parameter space, determined by relaxing the $M_Z$ low-scale
    constraint $f_1=\mzpred^2/\mzexp^2-1$.  Where the
    solid line crosses the horizontal 
    dashed line, the predicted value of $M_Z$ matches the experimental
    value ($f_1=0$) and we have a solution which satisfies all 11 low-scale
    constraints.
    We used $\tan\beta=40$, $A_0=0$~GeV, $m_0=2855$~GeV, $M_{1/2}=660$~GeV.}
\label{fig:muscan}
\end{center}
\end{figure}

The output of this line walking is a list of points in the 11-dimensional
high-scale parameter space, one point for each value of $V_5$.  These points
represent a function ($f_1=\mzpred^2/\mzexp^2-1$) defined on the high-scale
parameter space, and we plot all 11 projections of 
$f_1$ in Fig.~\ref{fig:muscan}, for both signs of $\mu$ (as illustrated
by Fig.~\ref{fig:muscan}d\footnote{Note that the sign of $\mu$ is an RGE
  invariant, therefore the $\sgn(\mu(\mgut))=\sgn(\mu(\msusy))$.}). Points
where the 
function $f_1$ crosses 0 correspond to full solutions of all boundary
conditions and RGEs. Fig.~\ref{fig:muscan}d shows that, for this CMSSM point,
there are two solutions for $\mu<0$ and one for $\mu>0$. These three solutions
are evident in each of the other ten projections, although sometimes they are
very close and hard to spot (for example in Fig.~\ref{fig:muscan}a). Each
individual projection is non-smooth, and none of the one dimensional
projections yield a uniquely defined function (i.e.\ for some values of each
abscissa coordinate, there is more than one predicted $f_1$ value). 
The non-smoothness yields practical difficulties when solving the system:
the Newton-Raphson solver uses derivatives and the finite difference
approximation to them becomes huge in the vicinity of a non-smooth piece. 
This can send the solver off to extreme and unphysical parts of parameter
space, where we cannot obtain sensible solutions of the RGEs (for example,
because of poles in renormalising quantities, or because tachyons are
predicted). In this case, the attempt in question must be abandoned, and
another started with a different random starting point. If that starting point
is on the correct side of the kink, such that it can reach a full solution
without encountering the kink again, then a solution may be found. This
illustrates one of the reasons for requiring a stochastic approach where we
fire many different shots in order to find the different solutions. 
To summarise: non-smoothness sometimes provide difficulties for the
Newton-Raphson solver, whereas (as explained in Section~\ref{sec:fpi}), the
magnitude 
of the derivatives at the horizontal zero line affects the stability
of the FPI algorithm.

\begin{figure}
\begin{center}
\includegraphics[width=0.49\textwidth]{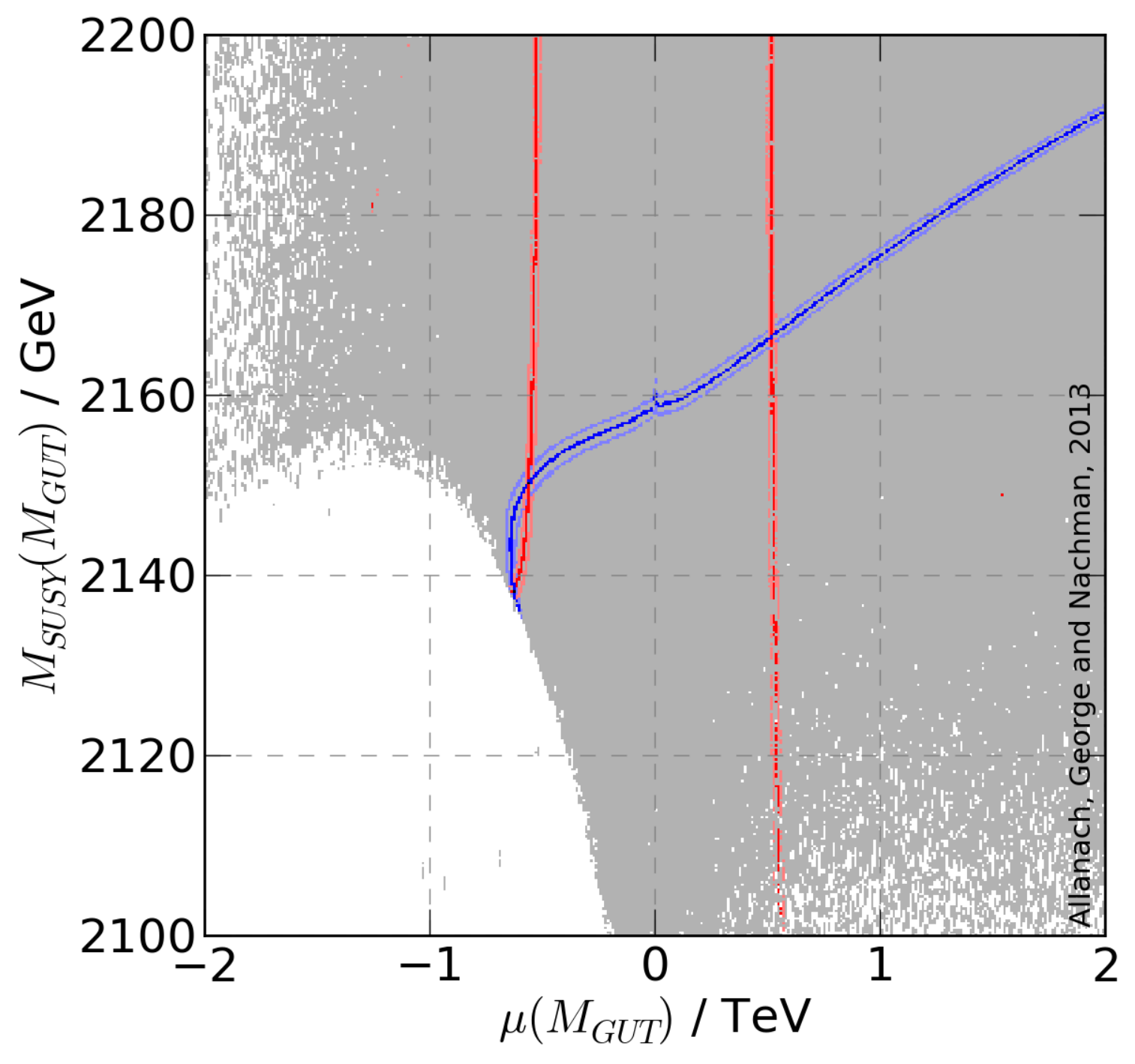}
\includegraphics[width=0.49\textwidth]{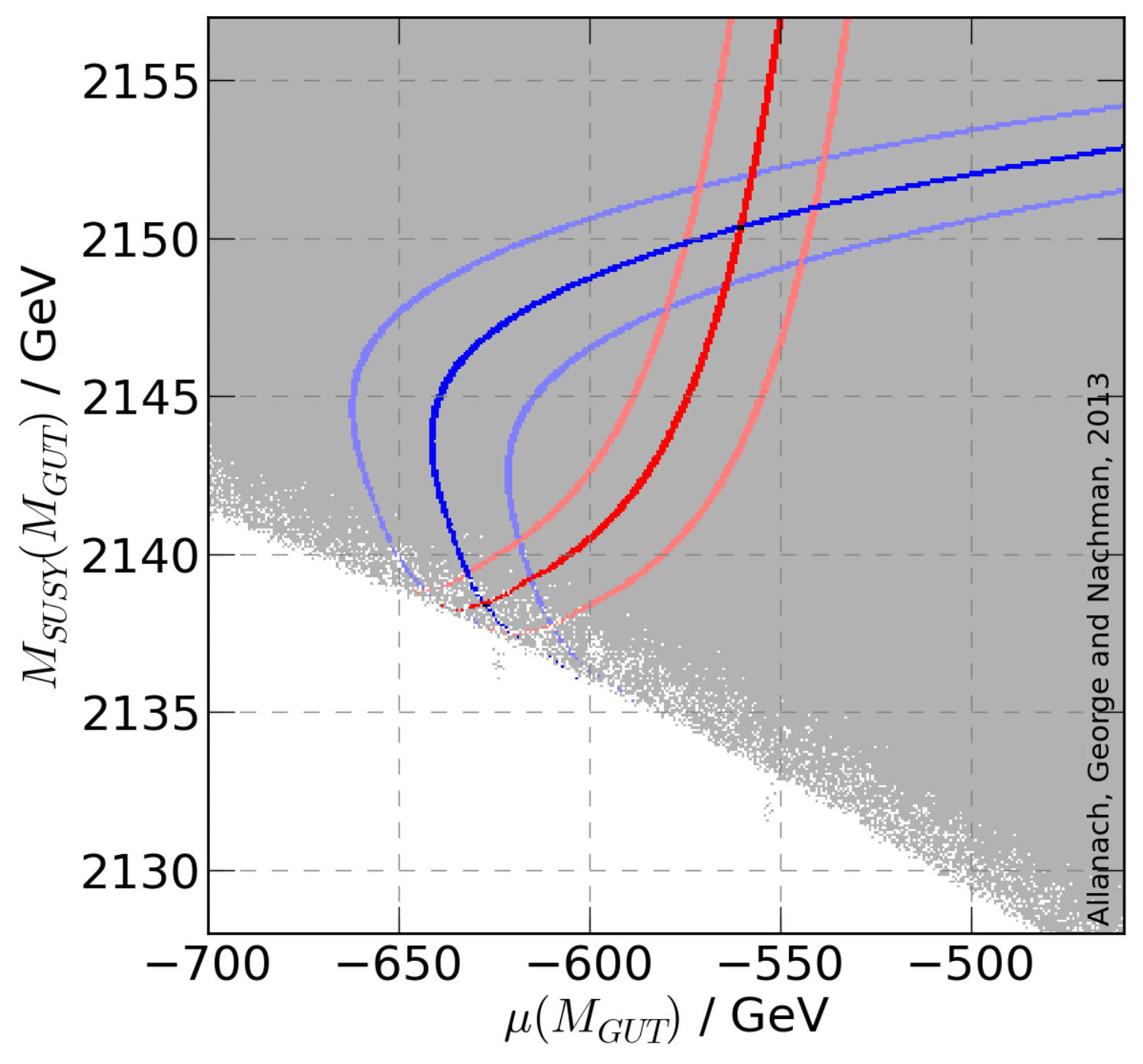}
\end{center}
\caption{Exploration of parameter space found by relaxing 
  $f_1$ and $f_{11}$ in the CMSSM with for $\tan\beta=40$, $A_0=0$~GeV,
  $m_0=2855$~GeV and $M_{1/2}=660$~GeV. 
    On the left we show a larger domain, but on the right we show a zoomed in
    domain of the $\mu<0$
    region including two solutions of the full system.
    On the centre red contour line, $M_Z^2$ is correctly predicted, on the
    centre blue contour, $\tan\beta(\mzexp)$ is.  Where these contours intersect, all
    11 constraints are satisfied; this occurs 3 times.  The additional red
    contours on either side of the centre contour show where the error
    $f_1=\mzpred^2/\mzexp^2-1$ is $+4$ and $-3$, i.e.\ the
    predicted $Z$ mass is 2 times the experimental one, physical or tachyonic,
    respectively.  For the additional blue contours $f_{11}$ is
    $\pm5\times10^{-3}$, 
    i.e.\ $\tan\beta(\mzexp)$ is 0.5\% higher or lower than the 
    input value, as per Eq.~\eqref{tanb}.
    The white region has tachyonic $A^0$s and does not
    yield physical solutions.}
\label{fig:rand9}
\end{figure}

In a separate study we then relaxed two constraints: $\mzpred^2$ and
$\tan\beta(\mzexp)$, equivalently $f_1$ and $f_{11}$.  We computed $f_1$
and $f_{11}$ over the two-dimensional plane of our freed-up
parameters, chosen here to be $\msusy(\mgut)$ and $\mu(\mgut)$, controlled by
$V_{11}$ and $V_5$ respectively.
Within a rectangular region in this plane we pick random initial points
and attempt to obtain a solution to the 9 low-scale constraints (we vary the
density of points near the full solutions to provide more clarity).
These functions each have contours at zero which predict the correct
value for a low-scale constraint, in direct analogy with the lines in
Fig.~\ref{fig:muscan} passing through zero.  Where these contours from
the two unconstrained $f$'s intersect, we have a complete solution satisfying all
11 constraints.
This is shown in Fig.~\ref{fig:rand9}. The standard solution for $\mu<0$ is at
higher values of $\msusy(\mgut)$. We can see from the figure that the additional
solution here is close to the physical boundary, where the $A^0$ becomes
tachyonic (shown as the white region in the plots). We also see from the figure that $\tan
\beta(\mzexp)$ varies much more slowly across parameter space than 
$\mzpred^2$. The figure also illustrates how difficult it is 
in general to obtain a solution: there is a large volume of unphysical (white)
parameter space, and indeed the outer side of the red contours the $Z^0$ is
tachyonic, yielding unphysical parameter space. The problem is much
exacerbated in the full 11-dimensional case, rather than just this
two-dimensional scan. 

\begin{figure}
\begin{center}
\includegraphics[width=0.6\textwidth]{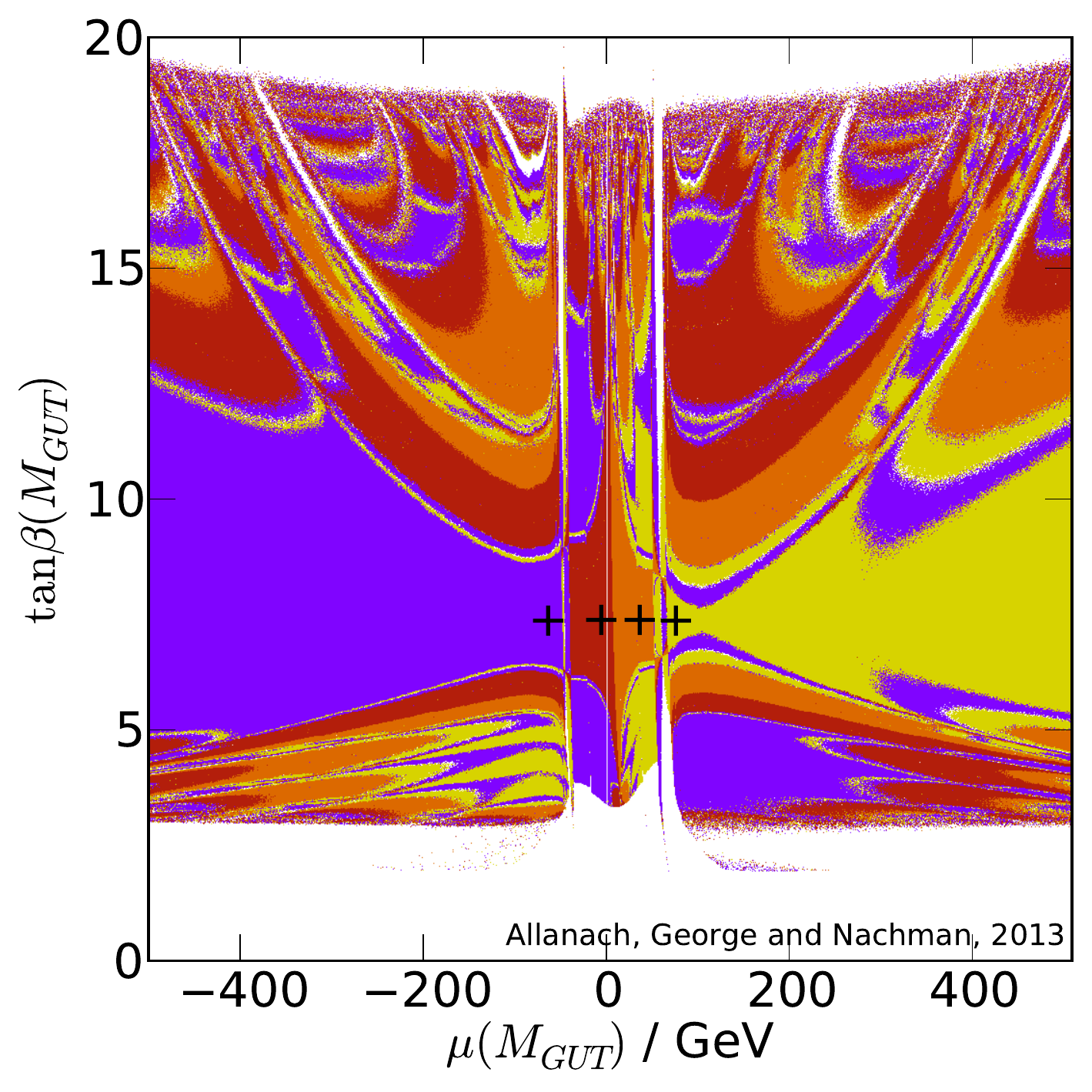}
\end{center}
\caption{Exploration of the high-scale parameter space to find the
    `catchment volume' of the Newton-Raphson solver in the $(\mu,\tan\beta)$
    plane.  Each point in the figure denotes a specific starting point for the
    high-scale parameters, and the colour of the point tells which
    solution the solver converges on.  There are 4 different solutions.
    The values of $\mu$ and $\tan\beta$ for the converged solutions are shown
    by the black plus-symbols.  White indicates that the solver could not
    converge 
    with the corresponding initial high-scale parameters.
    We used $\tan\beta=10$, $A_0=0$~GeV, $m_0=2800$~GeV, $M_{1/2}=200$~GeV.}
\label{fig:catchment}
\end{figure}

We also studied the behaviour of the Newton-Raphson solver by
determining the `catchment volumes' of the various solutions for one
particular point in CMSSM parameter 
space.  Figure~\ref{fig:catchment} shows the result.  We first find a solution
using the FPI algorithm, and use this to fix 9 of the
high-scale parameters.  The other 2, being $\mu(\mgut)$ and $\tan\beta(\mgut)$,
are scanned across a grid, as per the axis of the figure.  Each point then
corresponds to a distinct point in the high-scale parameter space, and we
then use it as the {\em initial}\/ guess for the Newton-Raphson solver.  
We then run
the solver until it converges (or not) on a solution.  If it converges, we
colour 
the point depending on which solution it finds.  In this example there are 4
different solutions, 2 with $\mu<0$ and 2 with $\mu>0$.  The results are very
sensitive to the particulars of the solver (such as tolerance, and criteria
for convergence), but they serve to illustrate the fact that there are many
sharp and non-trivial features in the equations that any solver must deal with. 

\section{Summary and conclusions}
\label{sec:summary}

We have provided a method to search in an efficient, multi-dimensional way for
multiple solutions to RGEs. The methodology that we employ should work in
general for theories which have multiple renormalising quantities, with
boundary conditions placed upon them at different renormalisation scales. 
It dispenses with fixed point iteration (the algorithm used by publicly
available spectrum calculators), which can have stability problems
for some of the solutions, and uses the shooting method instead. One shoots at
low-scale boundary conditions only in one direction, by first imposing high
scale boundary conditions and then running the RGEs down. Standard methods to
solve non-linear simultaneous equations are then used~---in this paper both
Broyden's method and Newton-Raphson~--- to find solutions to the low-scale
boundary conditions. To find several solutions, several shots are taken, each
with random starting values of the high-scale parameters consistent with the
high-scale boundary conditions. 
One problem that will be present in any such numerical scan is that one can
never be completely sure that one has found all of the solutions, unless
something analytically can be said about their number.

In a previous work~\cite{Allanach:2013cda} it was shown that the CMSSM 
has multiple solutions in some parts of its parameter space. The multiple
solutions are characterised by the fact that they have the same $m_0$, $M_{1/2}$,
$A_0$, $\tan\beta$ and $\sgn(\mu)$, yet are physically distinct,
with different sparticle masses and couplings. The existence of
multiple solutions was demonstrated
by inverting one of the boundary conditions and using fixed point iteration in
the other 10 dimensions of parameter space. Some solutions may be unstable to
fixed point iteration, however, which means that such an algorithm will never find
them. The shooting method has no such stability issues (that we are aware of),
and can in principle find all of the
solutions. This is done by randomly perturbing an initial guess of high-scale
parameters consistent with the high-scale boundary conditions, then homing in
on a solution with a non-linear multi-dimensional simultaneous equation
solver. One repeats this several times, hoping with each `shot' to find a
different solution.  Given enough shots, the method can find all of the
solutions. However, for a finite number of shots, one can never guarantee that
one has found all of them. 
In the present paper, we have shown that using the shooting method instead
of fixed point iteration
not only finds the previously discovered additional solutions, but also
finds new ones too. It was also our intention here to study the multiple
solutions phenomenologically, to see whether they are ruled out by
experimental data, and to see to which extent they differ from the standard
solutions (and if, for example, LHC searches need to be reinterpreted taking
the multiple solutions into account).

We find that CMSSM multiple solutions come in two classes. One class, at
extreme values of large $m_0$ just under the no electroweak symmetry breaking
limit, is ruled out by LEP2 constraints on chargino masses. The other class is 
potentially phenomenologically viable, however. Studying this latter class, we
see that sparticle masses are rather similar: most sparticles' masses are
within 1$\%$ of the standard solution's. The most constraining LHC searches
(various jets plus missing transverse momentum searches) are unlikely to have
a large difference between solutions: if a CMSSM point is ruled out by such an
analysis in the standard solution, it will be also ruled out for the
additional solution (and vice versa). 
However, chargino masses may be
around 10$\%$ different between solutions and so any CMSSM LHC analyses
relying on charginos in decay chains are vulnerable to correction coming from
the existence of the multiple solutions. Indeed, branching ratios of decays of
stops and sbottoms into charginos are shown to differ by several tens of
percent. 
Precision measurements at a future linear collider
facility could certainly be sensitive to percent or per-mille differences in
various sparticle masses, and so exclusion or measurement would need to take
the additional solutions into account. 

We predicted the thermal relic density of neutralinos, comparing the standard
and 
the new, phenomenologically viable class of solutions. Here, we see large
differences: 
a factor of 2--3 in $\Omega_\text{CDM} h^2$ is possible between the
predictions of the two separate solutions. We have illustrated with a point
that predicts far too much dark matter for the standard solution, where
nevertheless the additional solution is near the observational central value
derived by Planck.
Thus, we conclude that analyses involving $\Omega_\text{CDM} h^2$ as a
constraint 
should, of necessity, take multiple solutions into account. Many recent CMSSM
analyses fit $\Omega_\text{CDM}$ and other
data (see, for example~\cite{Allanach:2005kz,Buchmueller:2011ab,Balazs:2012qc,Cabrera:2012vu,Fowlie:2012im,Strege:2012bt}),
and these should be modified to include the additional solutions. It could be,
for instance, that a better best-fit point exists within the additional
solutions, meaning that the $\chi^2$ fit frequentist type analyses' 95$\%$
confidence level contours move significantly. 

\section*{Acknowledgements}

This work has been partially supported by STFC\@.
DG is funded by a Herchel Smith fellowship.
We would like to thank other members of the Cambridge SUSY Working Group
for discussions and helpful comments. 

\appendix

\section{Definition of Quantities in the Boundary Conditions}
\label{sec:defs}

Here, we define the quantities appearing in Table~\ref{tab:bcs}. Running
quantities without a `pred' label are obtained by running the
initial parameters down in renormalisation scale from $V_{11}$. 
The other quantities in $f_{1}$ and $f_2$
come from the electroweak symmetry breaking conditions, i.e.\ 
\begin{equation}
\mzpred^2 =
2 \left( \frac{m_{\bar{H}_1}^2(V_{11}) -  m_{\bar{H}_2}^2(V_{11}) \tan^2
  \beta(V_{11})}{\tan^2 \beta(V_{11}) - 1} -  \mu^2(V_{11})
\right) + \Pi_{ZZ}^T(V_{11}),
\label{mzpred} 
\end{equation}
where in practice, we use $\Pi_{ZZ}^T(V_{11}) = {\mzexp}^2 -
M_Z^2(V_{11})$ for the self energy of the $Z$ boson,
$M_Z^2(V_{11})=v^2(V_{11}) [g_2^2(V_{11}) + \frac{3}{5} g_1^2(V_{11})]/2$
and
\begin{eqnarray}
  \tan \beta^{\text{pred}}(V_{11}) = \tan \frac{1}{2} \left[ \sin^{-1} 
    \left( \frac{2 m_3^2(V_{11})}{m_{\bar H_1}^2(V_{11}) + m_{\bar
          H_2}^2(V_{11}) + 
      2 \mu^2(V_{11})}\right)
\right].
\end{eqnarray}
$m_{\bar{H}_i}^2=m_{H_i}^2-t_i/v_i$  are fixed by the soft SUSY
breaking mass parameters for the Higgs fields $m_{H_i}^2$, 
$i \in \{1,2\}$, as well as by the tadpole
contributions $t_i$ coming from loops. 
In $f_3$, $\msusy^{\text{pred}}=\sqrt{m_{{\tilde t}_1}
  (V_{11}) m_{{\tilde t}_2}(V_{11})}$. The MSSM parameters are then evolved to
$\mzexp$, where the $M_Z$ boundary conditions are applied:
in $f_{4,5,6}$, $Y_{t,b,\tau}^{\text{pred}}(\mzexp)$ are the top,
bottom and tau Yukawa couplings 
predicted by the central values of the top, bottom and tau pole masses,
respectively, once all of the SUSY contributions have been added to the
pole masses $m_{t, b, \tau}$~\cite{Pierce:1996zz}:
\begin{eqnarray}
Y_{t}^{\text{pred}}(\mzexp) &=& \frac{\sqrt{2} (m_{t} + \Re
  \Sigma_{t})}{v(\mzexp) \sin   \beta(\mzexp)}
, \nonumber \\
Y_{b,\tau}^{\text{pred}}(\mzexp) &=& \frac{\sqrt{2} (m_{b,\tau} + \Re
  \Sigma_{b,\tau})}{v(\mzexp) \cos \beta(\mzexp)},
\end{eqnarray}
where $\Re \Sigma_{t,b,\tau}$ are the real parts of the one-loop MSSM
contributions to the top, bottom and tau masses, respectively.
$g_3^\text{pred}(\mzexp)$ in $f_9$ is extracted from the central value
of $\alpha_s(M_Z)$ in 
the $\overline{MS}$ scheme by changing to the modified DRED scheme and adding
MSSM contributions and
$g_{1,2}^\text{pred}(\mzexp)$ in $f_{7,8}$ are the gauge couplings
predicted by the 
central values of $\alpha(\mzexp)$ and the Fermi constant
$G_F$ in the manner described in Appendix D of Ref.~\cite{Pierce:1996zz}.
$f_{10}$, contains the boundary condition on the Higgs VEV, where
\begin{equation}
  {v^\text{pred}}^2 = 4 \frac{{\mzexp}^2 +
    \Pi_{ZZ}^T(V_{11})}{g_2^2(V_{11}) + \frac{3}{5}g_1^2(V_{11})}
\end{equation}
and all Higgs vacuum expectation values (VEVs) are in the Feynman gauge (as
are the loop  corrections), and $v^2$ is the Pythagorean combination of the
two MSSM Higgs VEVs, i.e.\ $v_1^2+v_2^2$.
In $f_{11}$, $\tan \beta$
refers to the value that is input and specifies the CMSSM parameter point. 

\bibliographystyle{JHEP-2}
\bibliography{paper}

\end{document}